%% file: main.tex
\documentclass[lettersize,journal]{IEEEtran}
\input{setup.tex}
\usepackage{orcidlink} 
\newcommand{\nooutlineorcid}[1]{%
    {\hypersetup{hidelinks}\orcidlink{#1}}%
}

\newif\ifblind
\blindfalse

\begin{document}
\setlength{\topsep}{0pt} 
\setlength{\textfloatsep}{0pt} 

\title{Analytic Roof\/line Modeling and Energy Analysis of LULESH Proxy Application on Multi-Core Clusters
}

\ifblind
\author{Authors omitted for double-blind review process}
\else
\author{
Ayesha Afzal\nooutlineorcid{0000-0001-5061-0438},~\IEEEmembership{member,~IEEE,}
Georg Hager\nooutlineorcid{0000-0002-8723-2781} 
and
Gerhard Wellein\nooutlineorcid{0000-0001-7371-3026}
\IEEEcompsocitemizethanks{\IEEEcompsocthanksitem 
A. Afzal, G. Hager and G. Wellein are with the Erlangen National High Performance Computing Center (NHR@FAU), Germany and the Department of Computer Science at Friedrich-Alexander-Universit\"at Erlangen-N\"urnberg, Germany. \protect 
E-mail: \{ayesha.afzal, georg.hager, gerhard.wellein\}@fau.de
}
}



\fi

\maketitle

\begin{abstract}
We present a thorough performance and energy consumption analysis of the LULESH proxy application in its OpenMP and MPI variants on two different clusters based on Intel Ice Lake (ICL) and Sapphire Rapids (SPR) CPUs. We first study the strong scaling and power consumption characteristics of the six hot spot functions in the code on the node level, with a special focus on memory bandwidth utilization. 
We then proceed with the construction of a detailed Roof{}line performance model for each memory-bound hot spot, which we validate using hardware performance counter measurements. We also comment on the observed discrepancies between the analytical model and the observations. To discern the influence of the programming model from the influence of implementation of the code, we compare the performance of OpenMP and MPI based on problem size, examining if the underlying implementation is equivalent for large problems, and if differences in overheads are more significant at smaller problem sizes.
We also conduct an analysis of the power dissipation, energy to solution, and energy-delay product (EDP) of the hot spots, quantifying the influence of problem size, core and uncore clock frequency, and number of active cores per ccNUMA domain.  Relevant energy savings are only possible for memory-bound functions by using fewer cores per ccNUMA domain and/or reducing the core clock speed. A major issue is the very high extrapolated baseline power on both chips, which makes concurrency throttling less effective. In terms of energy-delay product (EDP), on SPR only memory-bound workloads offer lower EDP compared to Ice Lake. 
\end{abstract}

\begin{IEEEkeywords}
LULESH proxy application, Roofline model, performance, power dissipation, energy consumption, Ice Lake and Sapphire Rapids processors
\end{IEEEkeywords}

\section{Introduction and related work}
\IEEEPARstart{A}{s} the complexity of modern high-performance computing (HPC) systems continues to grow, achieving optimal performance while minimizing energy consumption has become a paramount challenge for applications in the computational science and engineering field. 
In this paper, we study the performance and energy of the \Ac{LULESH} proxy application \cite{KarlinLULESH2012} on multi-core clusters.
LULESH is a popular benchmark in the HPC community because of its computational and memory access patterns, which are representative of real-world applications.
However, any performance and energy optimization effort requires a thorough comprehension of the interplay between hardware architecture and application behavior. Therefore, this work uses the Roof{}line Model \cite{roofline:2009} to derive performance limits and computational intensities of LULESH on two different clusters based on Intel Ice Lake (ICL) and Sapphire Rapids (SPR) CPUs. 
Additionally, given the growing emphasis on energy-efficient computing in modern HPC systems, we extend our study to include an energy analysis.
Our work offers valuable insights into how LULESH application performance and energy usage can be balanced to achieve better resource efficiency. 

\subsection{Related work}
Previous studies on LULESH have mostly concentrated on performance evaluation \cite{Carothers2017, Malony2020, Copik2021} and optimization \cite{Liu2015, León2015, Singanaboina2024}, while some efforts have been made to incorporate energy efficiency analysis \cite{León2016, Wu2017}.
Karlin et al. \cite{Karlin2013} compared different implementations of LULESH to determine the benefits and drawbacks of various programming models for parallel computation in terms of programmer productivity, performance, and ease of applying optimizations.
Marques et al. \cite{Marques2017} demonstrated the usefulness of a cache-aware diagnostic Roof{}line Model analysis within Intel Advisor \cite{IntelAdvisor2024} by analyzing LULESH to identify the critical bottlenecks.
Nevertheless, to the best of the authors' knowledge, no prior effort has yet been made to describe the LULESH application using predictive white-box analytic performance models such as Roof{}line.
Additionally, a comprehensive study of the energy aspects of LULESH on contemporary multi-core clusters is still lacking.

\subsection{Contribution}
This paper makes the following relevant contributions, which, although currently presented only for Intel-based clusters, offer metrics and analysis techniques that are valuable in a larger context.
\begin{itemize}[\setlength\topsep{0pt}]
    \item We present an overview of performance and energy metrics of LULESH in MPI and OpenMP modes at the ccNUMA domain, node, and multi-node levels on two clusters featuring different generations of modern Intel server CPUs. This comparison underscores the need to balance performance and energy consumption when optimizing parallel applications like LULESH.
    \item We provide a predictive Roof{}line performance model for five of LULESH's hot spot functions for the first time and validate it using hardware performance counter measurements.
    \item Through kernel-based analysis, we demonstrate that analyzing power dissipation and energy consumption requires a clear distinction between memory-bound and non-memory-bound kernels, the latter being the target for in-core optimizations such as SIMD vectorization and the former offering lower energy-delay product (EDP) on SPR compared to ICL.
    \item Using the full application, we compare the performance of OpenMP and MPI across problem sizes to discern the influence of the programming model from the influence of code implementation.
    \item We further show that the minimization of energy-delay product and energy to solution are dominated by code scaling characteristics and chip idle power, making concurrency throttling and the uncore clock speed settings less effective while core clock speed settings play a more significant role in improving efficiency. 
\end{itemize}

\subsection{Overview}
This paper is organized as follows:
We first provide an overview of our experimental environment and methodology in Sect.~\ref{sec:environment}, followed by an introduction to the LULESH proxy application and its implementation details in Sect.~\ref{sec:lulesh}.
In Sect.~\ref{sec:kernelAnalysis} we concentrate on performance, power, and energy by carrying out a \emph{kernel-based} analysis on the node level and constructing a Roof{}line performance model in Sect.~\ref{sec:kernelAnalysis}.
Section~\ref{sec:fullapplication} extends the analysis to the full application for both single- and multi-node scenarios.
Finally, Sect.~\ref{sec:conclusion} summarizes the paper and gives an outlook to future work.

\section{Test bed and experimental methodology} \label{sec:environment}
Table~\ref{tab:systems} outlines the hardware and software configurations employed for our experiments.
We used the following two Intel-based InfiniBand (HDR-100) clusters:
\begin{enumerate}
    \item ClusterA\footnote{\ifblind
URL omitted for double-blind review process \else\url{https://hpc.fau.de/systems-services/documentation-instructions/clusters/fritz-cluster}\label{foot:Fritz}\fi} comprises two Intel Xeon Ice Lake (ICL) CPUs per node, each with 36 cores.
    \item ClusterB$^\textrm{1}$
	    comprises two Intel Xeon Sapphire Rapids (SPR) CPUs per node, each with 52 cores.
\end{enumerate}
Sub-NUMA Clustering was enabled on both systems, leading to a fundamental scaling unit (i.e., one ccNUMA domain) of half (i.e., 18 cores) of ClusterA's socket on and one-fourth (i.e., 13 cores) of ClusterB's socket.
All hardware prefetching mechanisms were enabled, and hyper-threading was disabled on both clusters.
Consecutive OpenMP threads and MPI processes were mapped to consecutive cores using the \CODE{likwid-perfctr} and \CODE{likwid-mpirun}~\cite{LIKWID} startup wrappers, respectively. 
We did not test the HBM variant of Sapphire Rapids CPUs as it was not generally accessible at the time of writing.

\input{figures/tab_systems.tex}

Unless otherwise specified, the core clock frequency was fixed to the base values of the respective CPUs via the Slurm batch scheduler (\CODE{cpu-freq=<min\_freq\_in\_kHz>-<max\_freq\_in\_kHz>:performance}), avoiding any power-saving behaviors that would typically lower the frequency.
Whenever applicable, the uncore and turbo clock frequencies of the hardware were set using the \code{likwid-setFrequencies} tool. 
The expected clock frequency settings were verified using \code{likwid-perfctr}.

The Intel VTune Profiler~\cite{IntelVTune} and the LIKWID tool suite~\cite{LIKWID, likwidweb} were employed for reading hardware performance and energy counter measurements and validating results where necessary.
The analyses from Likwid and Vtune yielded comparable values; thus, we focus on reporting the results from Likwid. The \texttt{likwid-powermeter} tool was utilized to obtain theoretical insights on turbo mode. Given the 1~kHz update frequency of RAPL counters on ICL and SPR, we excluded measurements shorter than a few milliseconds to ensure accuracy.
All node-level experiments were conducted on the same node to reduce RAPL measurement variation across nodes.
Before collecting measurements, a few warm-up time steps were performed to stabilize runtime and eliminate first-call overhead. 
To account for runtime variability, we repeated each code execution multiple times, and only statistically significant deviations were reported.
We present results for Intel MPI implementation, as node-level findings with OpenMPI implementation showed only minor differences.

\subsection{LULESH configuration settings}
This study used the first official release (version 2.0) of the LULESH proxy application\footnote{\Ac{LULESH} proxy application: \url{http://asc.llnl.gov/codes/proxy-apps/lulesh}} \cite{KarlinLULESH2012}, which, by default, calculates time constraints (courant and hydro) and determines the minimum time step required across domains.
This dynamic time step calculation incurs extra reductions (\code{MPI\_Allreduce} via the \code{MPI\_MIN} operation) during $\code{domain.cycle}-1$ iterations, which are only necessary to enforce time constraints.
The number of steps to solution scales to an arbitrary size based on an analytical equation.
Independent of the time step size, we used a fixed iteration count of 300 (\code{domain.cycle} in code and \code{-i} on command line).
The domain size was varied from $60^3$ to $350^3$ for the OpenMP version and from $60^3$ to $150^3$ per process for the MPI version (\code{numElem} in code and \code{-s} on command line).
In the MPI parallelization of LULESH using a weak scaling approach, the total problem size is defined as $p \times s^3$, where $p$ represents the number of MPI processes (required to be a cubic number), and $s$ denotes the domain size per process.
The domain size was varied from $60^3$ to $350^3$ for the OpenMP version and from $60^3$ to $150^3$ for the MPI version (\code{numElem} in code and \code{-s} on command line).
For larger domain sizes, the working sets were at least ten times larger than the \LLC of a single chip, preventing the working set from fitting into the available cache\footnote{In Ice Lake and Sapphire Rapids processors, the \LLC consists of a non-inclusive victim L3 cache and L2 caches.}.
Load imbalance, which is enabled by default, was disabled by setting the cost (\code{-c}) and balance (\code{-b}) command-line options to zero.
The number of regions is highly problem-dependent, and at a minimum, it must equal the number of distinct materials. Therefore, we used 11 distinct regions, as defined by default.
The message profile indicates long-distance point-to-point communication is used, involving the \code{MPI\_Isend, MPI\_Irecv, MPI\_Wait, MPI\_Waitall} sequence.
We consistently applied the \code{-O3} optimization and the widest supported SIMD instruction set, specifically \code{-xCORE-AVX512 -qopt-zmm-usage=high}, supported by the Intel architectures.
For data taking with \CODE{likwid-perfctr}, the LIKWID marker API was included in the code to allow separate measurement on each loop of the hot spot functions.  
 
\subsection{Observables for analysis} 
The performance metric used is the Figure of Merit (FOM), defined as the number of elements solved (z) per second (s).  
Alternative metrics, though ignored here, are the wall-clock time (complete run time) and the grind time (run time to update a single zone for one iteration).
We focus on metrics such as memory bandwidth utilization, overall instructions per work, power dissipation, energy-to-solution per work, and the energy-delay product for both CPU and DRAM. 
These metrics help quantify the impact of factors like problem size, core and uncore clock frequency, number of active cores per ccNUMA domain, and number of nodes.
The merit of normalizing metrics by work, rather than using raw metric, ensures comparability across problem sizes and weak scaling scenarios.

\subsection{Open-source dataset artifact}
 All information required to reproduce the data in this work is available at \url{https://github.com/RRZE-HPC/LULESH-AD} \footnote{Zenodo DOI: \url{https://doi.org/10.5281/zenodo.14056331}} \cite{LULESHAD:2024}.
Reproducibility initiative dependencies provided by the repository include a comprehensive description of all data sets and figures.
These descriptions include machine state files that document the hardware and software environments, scripts that explain the experimental design and methodology, modified code that incorporates LIKWID markers, and supporting data for experimental results.

\section{{L}ivermore {U}nstructured {L}agrangian {E}xplicit {S}hock {H}ydrodynamics ({LULESH})}\label{sec:lulesh}
\subsection{Introduction}
LULESH is a MPI- and OpenMP-parallel proxy application developed by Lawrence Livermore National Laboratory (LLNL) for shock hydrodynamic simulation on unstructured grids.
Originally, it was designed to simplify the complex structure of a real application while retaining the typical characteristics of computational workload encountered in real-world shock hydrodynamics problems.
It is a multi-phase code whose functions have various computational and communication requirements.
It employs an explicit time integration method and partitions the spatial problem domain into multiple volumetric elements of the mesh to discretely estimate the hydrodynamic equations.
Functions are called on a region-by-region basis to make the memory access patterns non-unit stride and to allow for configurable artificial load imbalance for evaluation purposes.
The mapping between materials and regions is crucial in LULESH, where multiple materials are mapped onto multiple regions (subsets of the mesh) of varying sizes, all modeling the same ideal gas material.
This uneven distribution of regions and different amounts of computation per grid point might cause a load imbalance.
LULESH models how the material deforms over time under shock and employs the Lagrangian approach, where the computational mesh moves with the material.
This approach allows for the explicit tracking of density, pressure, and internal energy.

\input{figures/tab_algoritm}
\subsection{Implementation}
The LULESH algorithm, outlined in Table~\ref{tab:Fig_LULESHAlgo}, is implemented in C++.
Although LULESH code uses a Cartesian mesh (more specifically, cubes), it mimics the unstructured, complicated hexahedral mesh geometry by employing indirection arrays and an unstructured data layout. 
Two stages are involved in updating the physical quantities: first, the \emph{nodal updates} at the hexahedra's corners, and then, the \emph{element updates} at each hexahedra's center.
Positions and velocities are stored by \emph{nodes}, while energy and pressure are stored by \emph{elements}.
The first step of nodal updates entails computing the nodal forces from the elemental contributions of volume force and stresses, which is the most computationally demanding step.
Then, it performs a diagnostic check for negative volumes, and each nodal force element-wise receives an hourglass contribution.
To calculate new nodal velocities and positions, it further computes the accelerations with appropriate symmetry boundary conditions. 
The first step of element updates entails the calculation of elemental kinematic values and new elemental and regional artificial viscosities.
After that, material properties are applied to each element, and the \acf{EOS} is evaluated. 
Then, material properties are applied to each element and the \acf{EOS} is evaluated. 
An outer loop over the regions and an inner loop over the elements in a region are used in the implementation.
The communication of ghost fields happens twice in the code. First, the communication happens for the positions exchange to ensure the same nodal values of the neighboring elements.
Second, viscosity gradients are exchanged following their computation.

\begin{figure*}
    \centering
    \subfloat[ Code variants]{\includegraphics[scale=0.47]{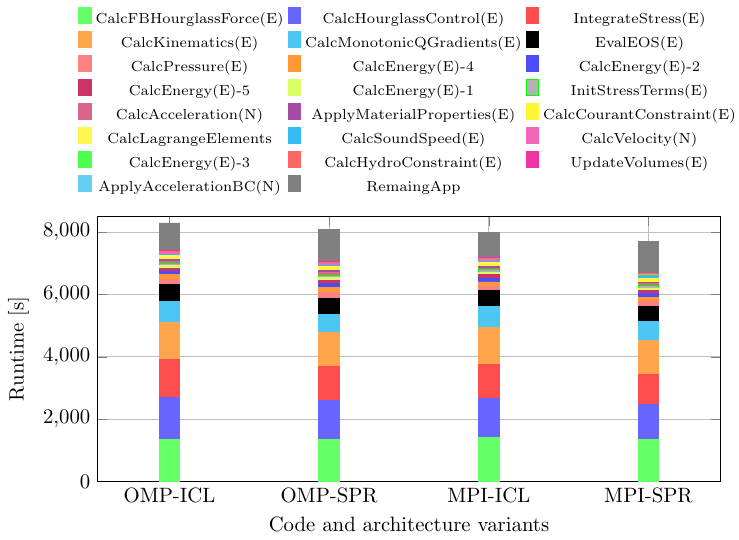}}\quad
    \subfloat[ Domain sizes: OMP-ICL variant]{\includegraphics[scale=0.47]{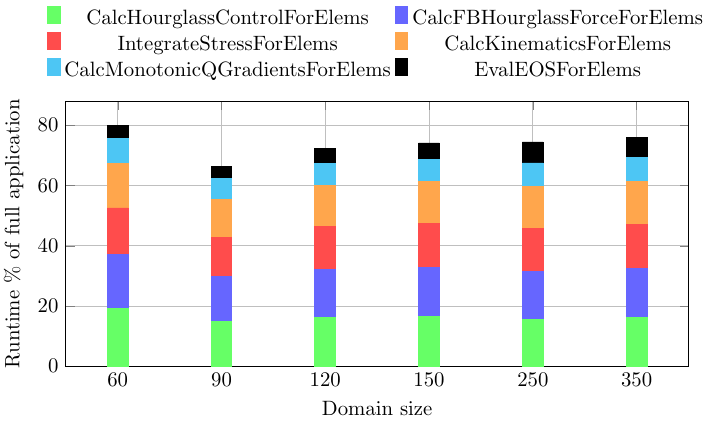}}\quad
    \subfloat[ Domain sizes: MPI-ICL variant]{\includegraphics[scale=0.47
]{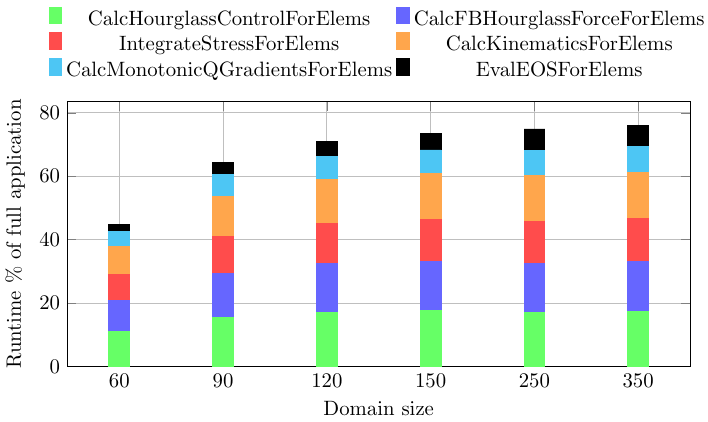}}
    \caption{Runtime breakdown of 22 LULESH functions on a single core for (a) four code variants at $350^3$ domain and (b, c) varying domain sizes for six hot spot functions in OMP-ICL and MPI-ICL versions.}
    \label{fig:22kernels-runtime}
\end{figure*}
\section{Breakdown analysis with profiling}\label{sec:kernelAnalysis}
This section offers basic insights into the performance, power, and energy-to-solution characteristics of hot spots within LULESH through profiling. We compare the OpenMP- and MPI-parallelized versions of the code on two clusters featuring Intel Ice Lake and Sapphire Rapids CPUs.

The focus of hot spot analysis is to examine the differences in performance and energy behavior between scalable and non-scalable functions. 
In addition, Section~\ref{sec:fullapplication} provides a comprehensive performance analysis for the full application, which includes a mix of memory-bound and non-memory-bound functions.

\subsection{Hot spot functions}
Most of the runtime of the application is in 22 compute kernels;  we instrumented these using the LIKWID marker API.
Among these, the following six are primary hot spot functions: 
\code{CalcHourglassControlForElems}, \code{CalcFBHourglassForceForElems}, \code{Integrate\-StressForElems}, \code{CalcKinematicsForElems}, \code{CalcMonotonicQGradientsForElems}, and \code{EvalEOSForElems}.

\begin{figure*}[t]
    \begin{minipage}{\textwidth}
        \centering
        \subfloat[ Bandwidth in OpenMP]
        {\includegraphics[scale=0.4]{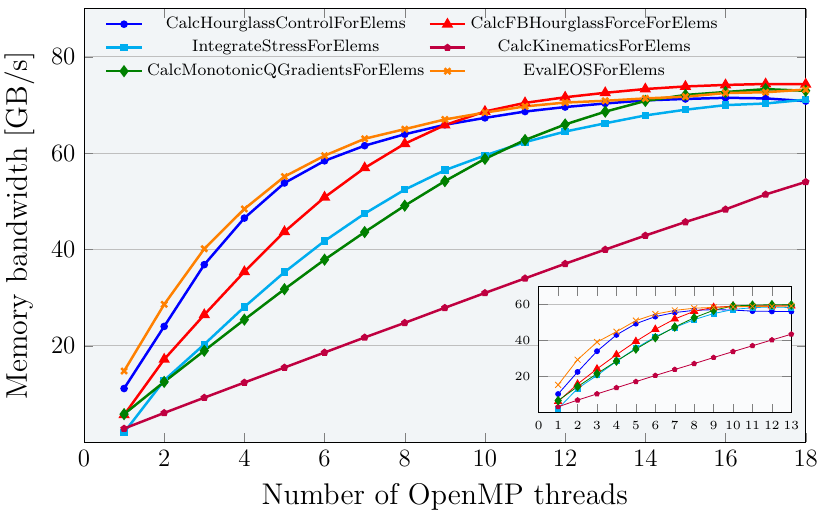}}\quad
        \subfloat[ Power in OpenMP] 
        {\includegraphics[scale=0.4]{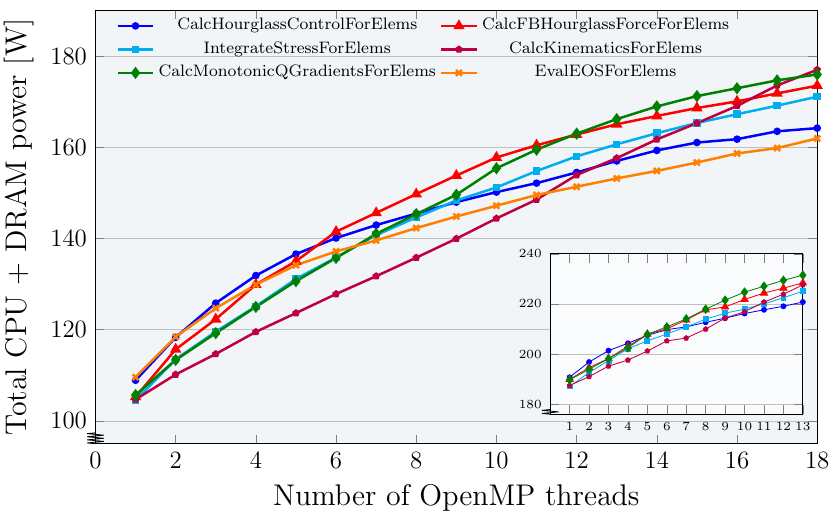}}\quad
        \subfloat[ Energy z-plot in OpenMP]{\includegraphics[scale=0.4]{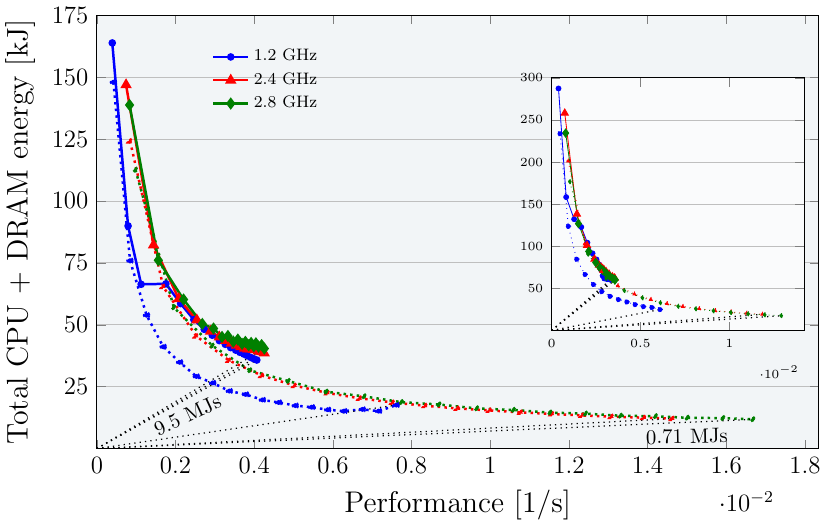}}
    \end{minipage}%
    
    \begin{minipage}{\textwidth}
        \centering
        \subfloat[ Bandwidth in MPI]
        {\includegraphics[scale=0.4]{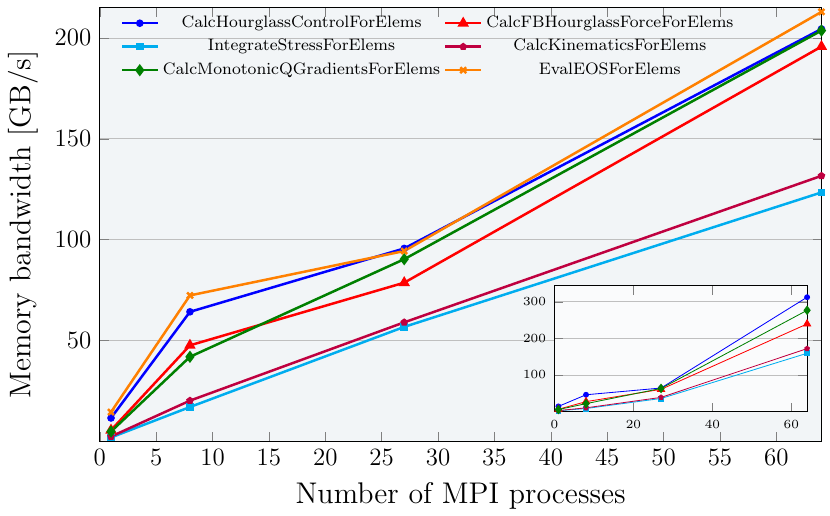}}\quad
        \subfloat[ Power in MPI]{\includegraphics[scale=0.4]{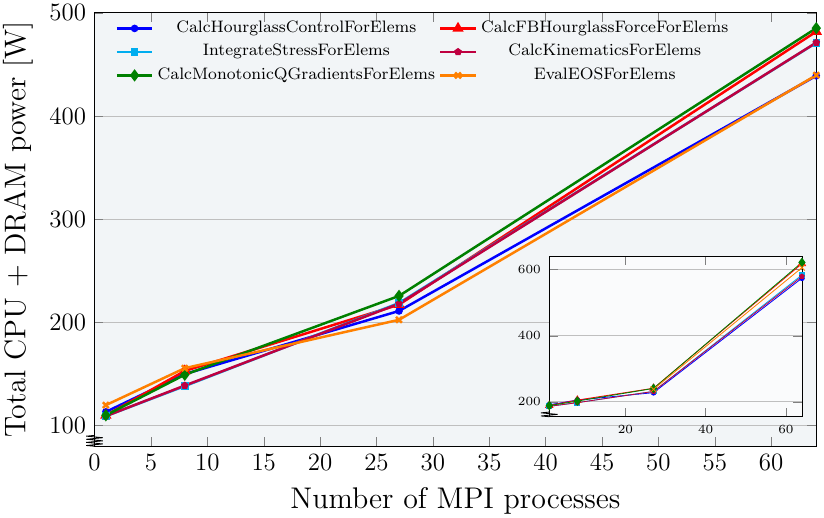}}\quad
        \subfloat[ Energy z-plot in MPI]{\includegraphics[scale=0.4]{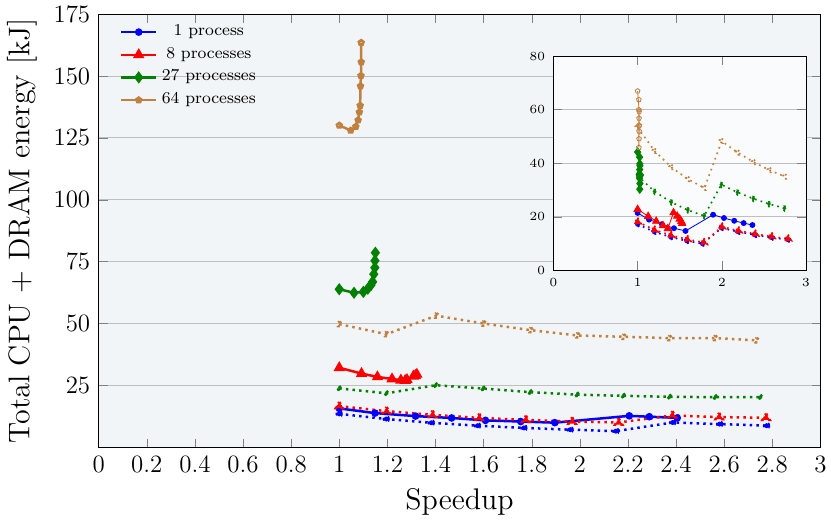}}
    \end{minipage}
    \caption{Performance-energy trade-offs for six LULESH hot spot functions, presented for both OpenMP (top) and MPI (bottom) versions on the ICL-based ClusterA node (subplot view in each plot: SPR-based ClusterB).
    The domain sizes have been selected as $350^3$ for OpenMP and $150^3$ for MPI.
    The plots illustrate: (a, d) bandwidth, (b, e) total power, and (c, f) a z-plot showing total energy-to-solution versus performance. The z-plot highlights the optimal EDP for different core counts and core frequencies (ranging from 1.0 to 2.8 \GHZ) at two corner cases: memory-bound hot spot function \code{CalcHourglassControlForElems} (solid) and non-memory-bound hot spot function \code{CalcKinematicsForElems} (dotted).}
    \label{fig:kernelAnalysis}
\end{figure*}

\subsection{Performance}
In Fig. \ref{fig:22kernels-runtime}(a), the 22 functions are listed in the legend in descending order based on their serial runtime at a problem size of $350^3$, with the remaining application runtime placed at the end.
For both OpenMP and MPI, the 22 kernels account for 90\% of total application runtime on ICL and 87\% on SPR. Among them, six primary hot spot functions contribute 76\% of the application's overall runtime on ICL and 73\% on SPR.
More specifically, six hot spot functions take the \{16.4,16.2,14.6,14.3,7.9,6.8\} percentage of overall application runtime in order on OMP-ICL version. However, the values for the other three variants are comparable, ranging from 6\% and 16\%.

Figures~\ref{fig:22kernels-runtime}(b, c) show the percentage of runtime relative to the full application for the six most relevant hot spots with varying problem size. For large problems, these fractions are very similar between MPI and OpenMP, while there are significant differences at smaller problem sizes.
The runtime fractions relative to the 22 functions for the six hot spot functions remains unchanged (consistently accounting for 90\%) across all problem sizes.

In order to discern memory-bound from non-memory-bound functions, we measured the memory bandwidth separately for the 22 hot spot kernels and present the data for the six most relevant hot spots in Figs.~\ref{fig:kernelAnalysis}(a, d).
In the OpenMP version, only one out of 22 functions, \code{CalcKinematicsForElems}, is non-memory-bound. In contrast, the MPI version has five non-memory-bound functions, including two hot spot functions (\code{CalcKinematicsForElems}, \code{IntegrateStressForElems}) and three non-hot spot functions (\code{CalcCourantConstraintForElems}, \code{CalcHydroConstraintForElems}, \code{UpdateVolumesForElems}).
Figure \ref{fig:kernelAnalysis}(a) also shows that a few functions exhibit significant superlinear scaling when going from 1 to 2 cores.
Although the full MPI application on a single core overall executes faster than any OpenMP version (OMP-ICL and OMP-SPR), the two non-memory-bound hot spot functions in the MPI version take slightly longer to execute.  
\smallskip \highlight{\emph{Upshot}: 
LULESH shows a mix of memory-bound and non-memory-bound hot spot functions, which behave in accordance with expectations. 
The hot spot function contributions are similar between OpenMP and MPI, though four memory-bound functions in OpenMP show scalable behavior in MPI, including one hot spot function.}

\subsection{Power, energy-to-solution and EDP}
Figure \ref{fig:kernelAnalysis}(b, e) shows that the total CPU and DRAM power dissipation of the hot spot functions (162--177~W for OMP-ICL on one ccNUMA domain and 439--485~W for MPI-ICL on 64 cores) eventually lies within the power range of the entire LULESH application (170~W for OMP-ICL and 451~W for MPI-ICL, which will be discussed later in Figure \ref{fig:fullAppAnalysis-energy}). When comparing memory-bound functions with the non-memory-bound function, the scalable function (purple) exhibits a distinct power pattern: it begins with low power at one core, aligning with other non-scalable functions, and then increases linearly as scaling progresses, ultimately reaching the highest power level among all functions across the full ccNUMA domain due to significant usage of memory bandwidth and in-core resources.
However, the non-scalable functions saturate more quickly with the number of cores, causing them to heat up faster.

In Fig. \ref{fig:kernelAnalysis}(c) we present a z-plot~\cite{AfzalThesis:2015,Wittmann:2016} with energy on the y-axis, performance on the x-axis, and the number of cores as the parameter along the data sets for the OpenMP version at a problem size of $350^3$. 
In a z-plot\footnote{The z-plot shows how energy consumption vs.\ code performance behaves when changing  the number of cores, the clock frequency, or any other relevant control parameter. It thus allows studying energy-performance trade-offs in a graphical way.}, horizontal and vertical lines indicate constant energy and performance, respectively. If the problem size is constant, each line through the origin is the locus of constant energy-delay product (EDP, the product of energy to solution and runtime), and its slope is proportional to the EDP. Optimizing for EDP is thus a search for the point in the z-plot that lies on the smallest-slope EDP line. In The data in For brevity, we present z-plot results for only three frequency settings: at, below, and above the base frequency, i.e., \{1.2, 2.4, 2.8\} \GHZ\ and for the two functions \code{CalcHourglassControlForElems} (solid) and \code{CalcKinematicsForElems} (dotted).
For comprehensive results on the impact of all frequencies on performance, power, and energy, we refer to Figures \ref{fig:App-A} and \ref{fig:App-B} of Appendix \ref{App:A}.
The six hot spot functions account for 74\% (672 kJ out of 905 kJ) of the total energy consumption of the entire application on a single core, and 58\% (133 kJ out of 231 kJ) on a single ccNUMA domain. The energy consumption of the scalable function (dotted line in Fig. \ref{fig:kernelAnalysis}(c)) initially aligns with that of the non-scalable function but decreases more sharply as scaling improves. Ultimately, across the full ccNUMA domain, it uses less than half the energy of the \code{CalcHourglassControlForElems} function (solid line). This is not surprising since the memory-bound function's performance saturates at some point; more cores deliver marginally more performance but still increase the power consumption.  

Fig.~\ref{fig:kernelAnalysis}(c) shows that the non-memory-bound hot spot function attains the lowest EDP with all cores active on the ccNUMA domain and running at a clock speed of 2.8\,\GHZ, clearly exhibiting a ``race-to-idle'' characteristic. The energy to solution is minimal at this point as well. For the memory-bound hot spot, the situation is not so clear; lowest clock speed clearly entails lowest energy to solution, but there is little variation in the EDP in the range between 1.2 and 2.4\,\GHZ, and due to the dominating baseline power all cores must be used. 
However, concurrency throttling (using fewer cores) for memory-bound code could become effective if non-compact pinning is used instead.
In memory-bound functions, playing with frequency does not impact performance very much, but it does make a difference with respect to energy. On the other hand, in compute-bound scalable kernels, all that matters is frequency and number of cores. There is a minimum energy point with respect to clock frequency which is usually far away from the performance maximum.
At frequencies below the base frequency, a kink appears at a certain core count, becoming more pronounced for memory-bound kernels and less so for non-memory-bound kernels and full applications, which include a mix of both kernel types. This is due to a peculiar behavior of the uncore frequency with changing number of cores at core frequencies below the base value; full data is available in the appendix.

In the subplot inserts of Sapphire Rapids, compared to Ice Lake, its significantly higher baseline power reduces the distinction between power trends of scalable and non-scalable hot spots and makes energy to solution slightly severe for memory-bound kernels. However, as the scale reaches a full node, the energy to solution becomes comparable to ICL, resulting in a higher EDP on SPR, as shown later in Figure \ref{fig:fullAppAnalysis-energy}.

In Fig. \ref{fig:kernelAnalysis}(f) we show a z-plot for the MPI version at a problem size of $150^3$ per process with clock frequency as a parameter along each data set. On the x-axis we choose speedup as a metric since the problem size changes for changing MPI process counts (1, 8, 27, 64). The data shows again the qualitative difference between memory-bound and non-memory-bound kernels. For memory-bound kernels, performance gains cease beyond 1.4 GHz if most of a socket is utilized; any frequency increase beyond this point merely results in wasted energy without further performance benefits.  

\smallskip \highlight{\emph{Upshot}: 
The six hot spots fall in the same power range as the full application, consuming up to 58\% of the overall energy to solution at large problem sizes. 
The best frequency setting for optimal EDP differs significantly between scalable and non-scalable hot spots, but the base frequency of the chip is a good overall compromise.}

\section{Analytic Performance modeling}\label{sec:RooflineModel}
In this section, we construct and validate memory traffic models for the memory-bound hot spots found in the LULESH application.

\subsection{Computational intensity}
The computational intensity $I$ of a loop is the ratio of the number of floating-point operations $N_{F}$ to the data transfers to and from main memory in bytes, calculated as follows:
\bq \label{eq:I}
    I = \dfrac{N_{F}}{V_{LD} + V_{ST}}~[\FB]\cma
\eq
where $V_{LD}$ and $V_{ST}$ are the load and store data volumes, respectively.
For LULESH, we compute the computational intensity for each loop and compare it with the values obtained from hardware performance counters.
For our calculations, we consider the flop or byte count for one iteration of the outer loop, which has a loop length of ``\code{numElem}'' (representing the amount of work per domain; \code{-s}).
For example, at a cubic domain of size 350, ``numElem'' is $350^3$.

The two primary data structures used in the hot spot functions of LULESH are 
\code{Real\_t (double)} for double-precision floating-point numbers and \code{Index\_t (int32\_t)} for 4-byte signed integers.
Since the compiler may optimize or modify arithmetic expressions, counting the number of FLOPs manually is not reliable. Instead, we 
use the more accurate FLOP count provided by tools like LIKWID. This ensures a more precise analysis, as LIKWID directly counts FLOPs at the hardware level, reflecting any optimizations that may be applied by the compiler.
Further, for the load and store volume calculation, we compare the manual prediction against the measured load and store traffic in memory using LIKWID. This comparison helps verify whether factors such as data reuse and write-allocate behavior play any role.
For a fair comparison, we use \code{-qopt-streaming-stores=never} to disable streaming stores.

\subsubsection{Common functionality among three hot spots}
Before diving into a detailed discussion of the five memory-bound hot spot functions, we begin by calculating the data transfers involved in the common functionality shared by the first, third, and fourth hot spot functions.
At the beginning of these three functions, \code{domain.nodelist(i)} retrieves node indices for each element (line 4 of Listing~\ref{algo:1}, line 3 of Listing~\ref{algo:7} and line 5 of Listing~\ref{algo:10}) and loads the corresponding nodal coordinates from global arrays into local arrays using the \code{CollectDomainNodesToElemNodes} function (line 5 of Listing~\ref{algo:1}, line 4 of  Listing~\ref{algo:7} and lines 6--9 of Listing~\ref{algo:10}).
The expression \code{domain.nodelist(i)} yields a 4-byte index of eight consecutive nodes, which gets assigned to \code{elemToNode}.
\begin{figure}[t] 
\centering
\begin{lstlisting}[
    style=small, 
    caption= The \code{CollectDomainNodesToElemNodes} function is\\ called in three hot spot functions., 
    label=algo:2,
]
static inline void CollectDomainNodesToElemNodes( Domain &domain, const Index_t* elemToNode, Real_t elemX[8], Real_t elemY[8], Real_t elemZ[8]) {
    Index_t nd0i = elemToNode[0];
    // similarly for nd1i-nd7i
    elemX[0] = domain.x(nd0i);
    // similarly for elemX[1]-elemX[7], elemY[0]-elemY[7] and elemZ[0]-elemZ[7]
}
\end{lstlisting}
\end{figure}
In Listing~\ref{algo:2}, the \code{CollectDomainNodesToElemNodes} function 
then loads eight 4-byte indices from main memory, resulting in a total load of 32 bytes (lines 2–3). 
Each set contains eight elements, each eight bytes long, leading to a total of 8 $\times$ 24 bytes for loading the positional coordinates (\code{domain.x}, \code{domain.y}, and \code{domain.z}) in lines 4–5.
The index values \{\code{nd0i}, \code{nd1i}, \code{nd2i}, \code{nd3i}, \code{nd4i}, \code{nd5i}, \code{nd6i}, \code{nd7i}\} and elements \{\code{elemX}, \code{elemY}, \code{elemZ}\}, are updated and reused for the next iteration. In the best case,  these data transfers can then come from cache.

\subsubsection{Function \code{CalcHourglassControlForElems}}
This first hot spot function, presented in Listing~\ref{algo:1}, calculates the standard hourglass force based on the element's geometry and deformation.
It processes elements within the domain, with data transfers involving 228-bytes loads 
(lines 4--5) as detailed above.
In line 6 of Listing~\ref{algo:1}, the \code{CalcElemVolumeDerivative} function calls the \code{VoluDer} function (Listing~\ref{algo:3}) eight times to compute the volume derivatives needed for calculating the volume force contributions to each element, which are then summed with those of its eight surrounding nodes.
Without considering compiler optimizations, the source code of this function has a total of 585 floating-point operations. However, the compiler performs substantial rearrangement of the arithmetic expressions, so that the total measured flop count is just 325 ($N_{F}$). We will use this number in the analysis.
Since the arrays \{\code{pfx}, \code{pfy}, \code{pfz}, \code{x1}, \code{y1}, and \code{z1}\} are declared within the loop (lines 2--3 of Listing~\ref{algo:1}), they incur no data traffic from main memory.
\begin{figure}[t] 
\centering
\vspace{-15pt}
\begin{lstlisting}[
    style=small, 
    caption=
 \code{CalcHourglassControlForElems} hot spot function., 
    label=algo:1,
]
for (Index_t i=0 ; i<numElem ; ++i){
    Real_t x1[8], y1[8], z1[8];
    Real_t pfx[8], pfy[8], pfz[8];
    Index_t* elemToNode = domain.nodelist(i);
    CollectDomainNodesToElemNodes(domain,elemToNode,x1,y1,z1);
    CalcElemVolumeDerivative(pfx, pfy, pfz, x1, y1, z1);
    for(Index_t ii=0;ii<8;++ii){
        Index_t jj=8*i+ii; 
        dvdx[jj] = pfx[ii];
        // similarly for dvdy[jj] and dvdz[jj] 
        x8n[jj] = x1[ii];
        // similarly for xy8n[jj] and z8n[jj]
    }
    determ[i] = domain.volo(i) * domain.v(i);
}
\end{lstlisting}
\end{figure}
\begin{figure}[t] 
\centering
\vspace{-20pt}
\begin{lstlisting}[
    style=small, 
    caption= The \code{VoluDer} function is called 8 times through line 6 of \\ the \code{CalcHourglassControlForElems} function., 
    label=algo:3,
]
static inline void VoluDer(Real_t* dvdx, const Real_t x0, /* similarly for dvdy, dvdz, x1-x5, y0-y5, z0-z5 */){
    const Real_t twelfth = Real_t(1.0) / Real_t(12.0);
    *dvdx = (y1+y2)*(z0+z1)-(y0+y1)*(z1+z2)+(y0+y4)*(z3+z4) -(y3+y4)*(z0+z4)-(y2+y5)*(z3+z5)+(y3+y5)*(z2+z5);
    *dvdx *= twelfth;
    // similarly for *dvdy and *dvdz
}
\end{lstlisting}
\end{figure}
In lines 7--13 of Listing~\ref{algo:1}, each iteration of the outer \enquote{for-loop} involves six arrays \{\code{dvdx}, \code{dvdy}, \code{dvdz}\} and \{\code{x8n}, \code{y8n}, \code{z8n}\}, each of length eight times the number of elements.
These arrays are used to temporarily store volume derivatives and coordinates, respectively, which are later passed to the subsequent \code{CalcFBHourglassForceForElems} function.
This results in a data transfer of 8 $\times$ 48 bytes each for stores and loads due to write allocation.
Additionally, line 14 of Listing~\ref{algo:1} requires a data transfer of 8 bytes to store data for \code{determ[]} and 24 bytes for loading data for three arrays (\code{domain.volo}, \code{domain.v} and \code{determ[]}).
In the parallel multi-threaded case on a ccNUMA domain, the intensities slightly worsen due to marginally higher memory traffic.
\input{figures/Kernels/tab_Roofline}
The predicted and measured computational intensities, presented in Table \ref{tab:Roofline}, are nearly similar.
The slight discrepancy in measured memory traffic between SPR and ICL is likely due to the cache reuse for certain arrays.

\begin{figure}[t] 
\centering
\vspace{-15pt}
\begin{lstlisting}[
    style=small, 
    caption= \code{CalcFBHourglassForceForElems} hot spot function., 
    label=algo:4,
]
// declarations and initialization for gamma[][]
for(Index_t i2=0;i2<numElem;++i2){
    Real_t *fx_local,*fy_local,*fz_local,hgfx[8],hgfy[8], hgfz[8],coefficient,hourgam[8][4],xd1[8],yd1[8],zd1[8], ss1,mass1,volume13, volinv = Real_t(1.0)/determ[i2];
    const Index_t *elemToNode = domain.nodelist(i2);
    Index_t i3=8*i2;
    for(Index_t i1=0;i1<4;++i1){
        Real_t hourmodx = x8n[i3]*gamma[i1][0] +x8n[i3+1]*gamma[i1][1]+x8n[i3+2]*gamma[i1][2] +x8n[i3+3]*gamma[i1][3]+x8n[i3+4]*gamma[i1][4] +x8n[i3+5]*gamma[i1][5]+x8n[i3+6]*gamma[i1][6] +x8n[i3+7]*gamma[i1][7];
        // similarly for hourmody and hourmodz          
        hourgam[0][i1]=gamma[i1][0]-volinv*(dvdx[i3]* hourmodx+dvdy[i3]*hourmody+dvdz[i3]*hourmodz);
        // similarly for hourgam[1][i1]-hourgam[7][i1]
    }
    ss1 = domain.ss(i2);
    mass1 = domain.elemMass(i2);
    volume13 = CBRT(determ[i2]);
    Index_t n0si2 = elemToNode[0];
    // similarly for n1si2-n7si2
    xd1[0] = domain.xd(n0si2);
    // similarly for xd1[1]-xd1[7], yd1[] and zd1[]
    coefficient = - hourg*Real_t(0.01)*ss1*mass1/volume13;
    CalcElemFBHourglassForce(xd1, yd1, zd1, hourgam, coefficient, hgfx, hgfy, hgfz);
    if (numthreads > 1) {
        fx_local = &fx_elem[i3];
        fx_local[0] = hgfx[0];
        // similarly till fx_local[7], fy_local[], fz_local[]
    }
}
\end{lstlisting}
\end{figure}

\subsubsection{Function \code{CalcFBHourglassForceForElems}}
Listing~\ref{algo:4} presents the implementation of this second hot spot function, which calculates the Flanagan-Belytschko anti-hourglass force.
It executes 824 FLOPs, significantly contributing to the total FLOPS count of the application.
The index values can be reused, resulting in no data traffic from the main memory for loading indices for nodes in line 4, for each element's sound speed (\code{domain.ss()}) and mass (\code{domain.elemMass}) in lines 12--13, and for the eight consecutive nodes of the element in lines 15--16.
In lines 6--11, within the inner \texttt{for}-loop iterating four times, the 32-element \code{gamma[][]} array is locally declared, causing negligible impact on main memory traffic.
The \code{\{hourmodx, hourmody, hourmodz\}} are computed from the \code{\{x8n, y8n, z8n\}} arrays in lines 7--8 and stored in registers, while the hourglass modes \code{hourgram[][]} are calculated by loading the volume derivative arrays \code{\{dvdx, dvdy, and dvdz\}} in lines 9--10.
This loading process results in a total of 8 $\times$ 48 bytes being loaded from memory.
In line 14, the array determ[] (computed in the first hot spot) is retrieved from main memory with an 8-byte load each.
In lines 17--18, data transfer involves loading 192 bytes for nodal velocities.
Line 20 of Listing~\ref{algo:4} calls the \code{CalcElemFBHourglassForce} function, performing a significant amount of flops and incurring no data traffic from main memory as all arrays are declared within the loop (line 3). 
In lines 21--25, for the threaded version, the global values of anti-hourglass force \code{f*\_elem} are copied to local arrays \code{f*\_local} for each element to avoid race conditions. This copying results in a total of 8 $\times$ 24 bytes from the global array loaded from main memory and 8 $\times$ 24 bytes stored to the local array. 
These 192 bytes, each for load and store loads, are not included in the single-threaded predicted intensity values shown in Table \ref{tab:Roofline}. 
The intensity values measured with Likwid on single and multiple threads exhibit only a minimal variation with the calculated results.

\begin{figure}[t] 
\centering
\begin{lstlisting}[
    style=small, 
    caption= \code{IntegrateStressForElems} hot spot function., 
    label=algo:7,
]
for( Index_t k=0 ; k<numElem ; ++k ){
    Real_t B[3][8], x_local[8], y_local[8], z_local[8];
    const Index_t* const elemToNode = domain.nodelist(k);
    CollectDomainNodesToElemNodes(domain, elemToNode, x_local, y_local, z_local);
    CalcElemShapeFunctionDerivatives(x_local, y_local, z_local, B, &determ[k]);
    CalcElemNodeNormals(B[0], B[1], B[2], x_local ,y_local, z_local);
    if (numthreads > 1) {
     SumElemStressesToNodeForces(B,sigxx[k],sigyy[k], sigzz[k],&fx_elem[k*8],&fy_elem[k*8],&fz_elem[k*8]);
    }
}
\end{lstlisting}
\end{figure}

\begin{figure}[t] 
\centering
\vspace{-16pt}
\begin{lstlisting}[
    style=small, 
    caption= The \code{CalcElemShapeFunctionDerivatives} function\\ is called through line 5 of the \code{IntegrateStressForElems} function., 
    label=algo:8,
]
static inline void CalcElemShapeFunctionDerivatives( Real_t const x[], Real_t const y[], Real_t const z[], Real_t b[][8], Real_t* const volume ) {
    const Real_t x0 = x[0];
    // similarly for x1-x7, y0-y7 and z0-z7
    Real_t fjxxi, fjxet, fjxze, cjxxi, cjxet, cjxze; 
    // similarly for fjyxi, fjyet, fjyze and fjzxi, fjzet, fjzze and cjyxi, cjyet, cjyze and cjzxi, cjzet, cjzze
    fjxxi = Real_t(.125)*((x6-x0)+(x5-x3)-(x7-x1)-(x4-x2));
    fjxet = Real_t(.125)*((x6-x0)-(x5-x3)+(x7-x1)-(x4-x2));
    fjxze = Real_t(.125)*((x6-x0)+(x5-x3)+(x7-x1)+(x4-x2));
    // similarly for fjyxi, fjyet, fjyze, fjzxi, fjzet, fjzze
    cjxxi = (fjyet*fjzze)-(fjzet*fjyze);
    cjxet = -(fjyxi*fjzze)+(fjzxi*fjyze);
    cjxze = (fjyxi*fjzet)-(fjzxi*fjyet);
    // similarly for cjyxi, cjyet, cjyze, cjzxi, cjzet, cjzze
    b[0][0] = - cjxxi - cjxet - cjxze;
    b[0][1] = cjxxi - cjxet - cjxze;
    b[0][2] = cjxxi + cjxet - cjxze;
    b[0][3] = - cjxxi + cjxet - cjxze;
    b[0][4] = -b[0][2];
    b[0][5] = -b[0][3];
    b[0][6] = -b[0][0];
    b[0][7] = -b[0][1];
    // similarly for partials b[1][] and b[2][]
    *volume=Real_t(8.)*(fjxet*cjxet+fjyet*cjyet+fjzet*cjzet);
}
\end{lstlisting}
\end{figure}
\begin{figure}[t] 
\centering
\begin{lstlisting}[
    style=small, 
    caption= The \code{SumElemFaceNormal} function is called six times \\ through line 6 of the \code{IntegrateStressForElems} function., 
    label=algo:9,
]
static inline void SumElemFaceNormal(Real_t *normalX0, ... , Real_t *normalX3, const Real_t x0, ... , const Real_t x3, /* similarly for *normalY0-3, *normalZ0-3, y0-y3, z0-z3 */){
    Real_t bisectX0 = Real_t(0.5) * (x3 + x2 - x1 - x0);
    Real_t bisectX1 = Real_t(0.5) * (x2 + x1 - x3 - x0);
    Real_t areaX = Real_t(0.25) * (bisectY0*bisectZ1 - bisectZ0*bisectY1);
    *normalX0 += areaX;
    // similarly till *normalX3, and for bisectY, areaY, *normalY, bisectZ, areaZ, *normalZ 
}
\end{lstlisting}
\end{figure}
\subsubsection{Function \code{IntegrateStressForElems}}
The third hot spot function, whose implementation is shown in Listing~\ref{algo:7}, computes the stress tensor for each element
, with a total of 386 FLOPS.
As previously detailed in the common functionality, lines 3--4 involve no data transfers due to data reuse.
In line 5, \code{CalcElemShapeFunctionDerivatives} function (Listing~\ref{algo:8}) performs volume (Jacobian's determinant) calculations, involving extra work for numerical consistency. 
The function calls the 2D shape function derivatives \code{B[][]} and elements, declared locally in line 2 of Listing~\ref{algo:7} for each outer \enquote{for-loop} iteration, excluding data transfer from main memory by redeclaring them for each iteration.
Data transfer occurs only when the value is stored in the volume at line 23 of Listing~\ref{algo:8}, resulting in a transfer of 8-byte stores.
In line 6 of Listing~\ref{algo:7}, the \code{CalcElemNodeNormals} function calls the \code{SumElemFaceNormal} function (Listing~\ref{algo:9}) six times to determine the face normal of the element.
This normal computation results in a 6 $\times$ 24-byte load from main memory.
Finally, in lines 7--9 of Listing~\ref{algo:7}, the threaded version avoids thread writing conflicts at the nodes by providing each element with its own copy for writing.
This results in a total of 8 $\times$ 24 bytes loaded from main memory and 8 $\times$ 24 bytes stored for \code{fx[i]}, \code{fx[i]} and \code{fz[i]}, with i iterating 8 times. 
Table \ref{tab:Roofline} presents the calculated and measured computational intensity values for this third function, revealing only minor variance between them.

\subsubsection{Function \code{CalcMonotonicQGradientsForElems}}
Listing~\ref{algo:10} shows the implementation of this fourth hot spot that computes the gradients for artificial viscosity, called ``monotonic Q.''
It performs a total of 205 FLOPS, with the main bottleneck in runtime being the square root and division operations.
As noted earlier in the common functionality, lines 5--9 do not involve data transfers due to data reuse, omitting the reloading of indices \code{{i, n}} and position coordinates \code{{x, y, z}} from the main memory, which were already loaded in previous functions. 
In line 10, data transfer consists of loading 8 $\times$ 24 bytes for nodal velocities \code{{xv, yv, zv}}.  
The reference volume for each element is accessed using the domain class elements \code{domain.volo(i)} and \code{domain.vnew(i)} (line 11).
In lines 15--25, other domain class elements (\code{delv\_zeta, delv\_eta, delv\_xi, delx\_zeta, delx\_eta, delx\_xi}) involve a 2 $\times$ 48-byte data transfer for both load and store operations, followed by an 8-byte store in line 26.
\begin{figure}[t] 
\centering
\vspace{-15pt}
\begin{lstlisting}[
    style=small, 
    caption= \code{CalcMonotonicQGradientsForElems} hot spot function., 
    label=algo:10,
]
for (Index_t i = 0 ; i < numElem ; ++i ) {
    const Real_t ptiny = Real_t(1.e-36);
    Real_t ax,ay,az;
    Real_t dxv,dyv,dzv;
    const Index_t *elemToNode = domain.nodelist(i);
    Index_t n0 = elemToNode[0];
    // similarly for n1-n7
    Real_t x0 = domain.x(n0) ;
    // similarly for positional coordinates x1-x7, y, z 
    // similarly for nodal velocities xv, yv, zv
    Real_t vol = domain.volo(i)*domain.vnew(i) ;
    Real_t norm = Real_t(1.0) / ( vol + ptiny ) ;
    Real_t dxj = Real_t(-0.25)*((x0+x1+x5+x4)-(x3+x2+x6+x7));
    // similarly for dyj, dzj, dxi, dyi, dzi, dxk, dyk, dzk 
    /* find delxi and delvi (j cross k) */
        ax = dyi*dzj - dzi*dyj ;
        // similarly for ay and az
        domain.delx_zeta(i)=vol/SQRT(ax*ax+ay*ay+az*az+ptiny);
        ax *= norm ;
        // similarly for ay and az
        dxv=Real_t(0.25)*((xv4+xv5+xv6+xv7)-(xv0+xv1+xv2+xv3));
        // similarly for dyv and dzv
        domain.delv_zeta(i) = ax*dxv + ay*dyv + az*dzv;
    // similarly find delxi and delvi (j cross k)
    // similarly find delxj and delvj (k cross i)
    domain.delv_zeta(i) = ax*dxv + ay*dyv + az*dzv ;
}
\end{lstlisting}
\end{figure}
The computational intensities measured using Likwid show minimal variance from the analytical predictions, confirming data reuse, as presented in Table \ref{tab:Roofline}.
\subsubsection{Function \code{EvalEOSForElems}}
Listing~\ref{algo:11} presents the implementation of the final hot spot function that calculates the pressure for each element using the \acf{EOS}.
Given the regions are evenly distributed, \code{numElemReg} is the total domain size divided by 11 regions.
In the first loop (lines 2--5), five double words (40 bytes) are loaded into the domain class's elements, which return energy \code{e}, pressure \code{p}, artificial viscosity \code{q} and linear and quadratic coefficients \code{ql, qq}.
In line 3, since the relative volume \code{delv} was already loaded in the previous \code{CalcKinematicsForElems} function, it can be reused without incurring additional data transfers from main memory. 
In the second loop (lines 8--12), 48 floating-point operations are performed, with 16 bytes of data transferred for stores and 8 bytes for loading \code{vnewc}.
The subsequent section of the code (line 14) includes conditional statements evaluating the Equation of State, which incur no data traffic.
The computational intensity values measured using Likwid closely match the predicted values, as shown in Table \ref{tab:Roofline}.
\smallskip \highlight{\emph{Upshot}: The measured operational intensities across the hot spots closely align with the analytical predictions.}
\begin{figure}[t] 
\centering
\begin{lstlisting}[
    style=small, 
    caption= \code{EvalEOSForElems} hot spot function., 
    label=algo:11,
]
for (Index_t i=0; i<numElemReg; ++i) {
    Index_t ielem = regElemList[i];
    delvc[i] = domain.delv(ielem);
    e_old[i] = domain.e(ielem);
    // similarly for p_old[], q_old[], qq_old[] and ql_old[]
}
for (Index_t i = 0; i < numElemReg ; ++i) {
    Index_t ielem = regElemList[i];
    Real_t vchalf;
    compression[i] = Real_t(1.) / vnewc[ielem] - Real_t(1.);
    vchalf = vnewc[ielem] - delvc[i] * Real_t(.5);
    compHalfStep[i] = Real_t(1.) / vchalf - Real_t(1.);
}
// EOS evaluating conditional statements ommited for brevity
\end{lstlisting}
\end{figure}

\subsection{Roof{}line Modeling}
The Roof{}line model ($P = \min (P_\mathrm{peak}, I\times b_s)$) analytically quantifies the hardware-software interaction by relating peak performance ($P_\mathrm{peak}$) and memory bandwidth of hardware ($b_s$) to computational intensity ($I$) of a loop.
The performance is limited by either $P_\mathrm{peak}$ or $I\times b_s$, indicating compute- or memory-boundedness, respectively.

\begin{figure}[t]
    \centering
    \subfloat[OMP-ICL]{\includegraphics[scale=0.42]{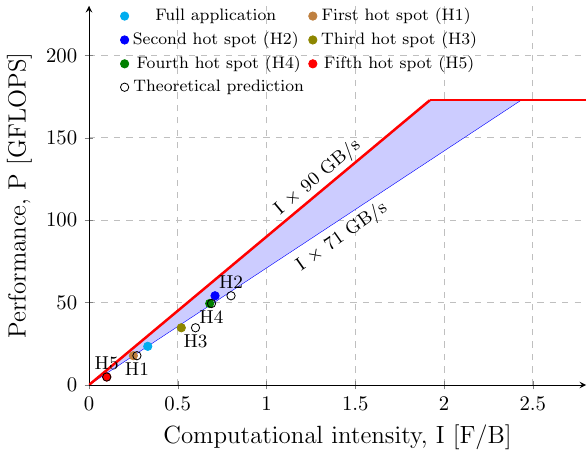}}\quad
    \subfloat[OMP-SPR]{\includegraphics[scale=0.42]{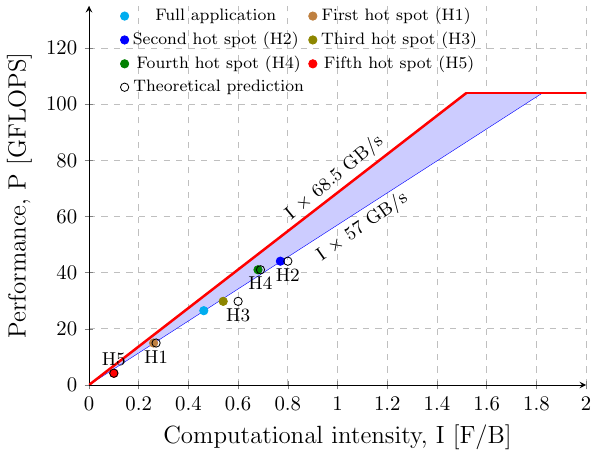}}
    \caption{The Roofline model for the OpenMP-parallelized LULESH application is shown, presenting the measured values for both the full application and individual hot spots (filled circles) against analytical predictions (empty circles).
    This data is presented for a single ccNUMA domain on two systems: (a) Ice Lake and (b) Sapphire Rapids. 
    }
    \label{fig:Roofline}
    \vspace{1em}
\end{figure}

In Figure \ref{fig:Roofline}, two lines are plotted to represent bandwidth, with a band between the minimum and maximum values.
A single ccNUMA domain of the Ice Lake (Sapphire Rapids) system has a theoretical memory bandwidth of 102.4\,\GBS\ (76.8\,\GBS), with the maximum achievable read-only bandwidth of 90\,\GBS\ (68.5\,\GBS) reflecting the most favorable benchmark conditions, while both the update benchmark and LULESH proxy application reach a maximum of 71\,\GBS\ (57\,\GBS). 
Since LULESH is not completely vectorized, the peak performance ceiling of the model is calculated using the scalar limit.
The theoretical scalar peak performance for one ccNUMA domain is 172.8\,\GFS\ on the Ice Lake system and 104\,\GFS\ on the Sapphire Rapids system.
Measured performance is plotted since predicted performance always aligns with the bandwidth ceiling, making it unnecessary to plot as a data point.
Predicted values are denoted by empty circles, while the filled color circle denotes the measured values obtained from LIKWID.
Finding indicates that intensities are fairly predicted and highlights the interaction between computation and memory bandwidth using the Roof{}line model, where programming model overhead appears as a gap between measured performance and the Roof{}line limit. 

\begin{figure*}[t]
    \begin{minipage}{\textwidth}
        \centering
        \subfloat[ Bandwidth in OpenMP]{\includegraphics[scale=0.4]{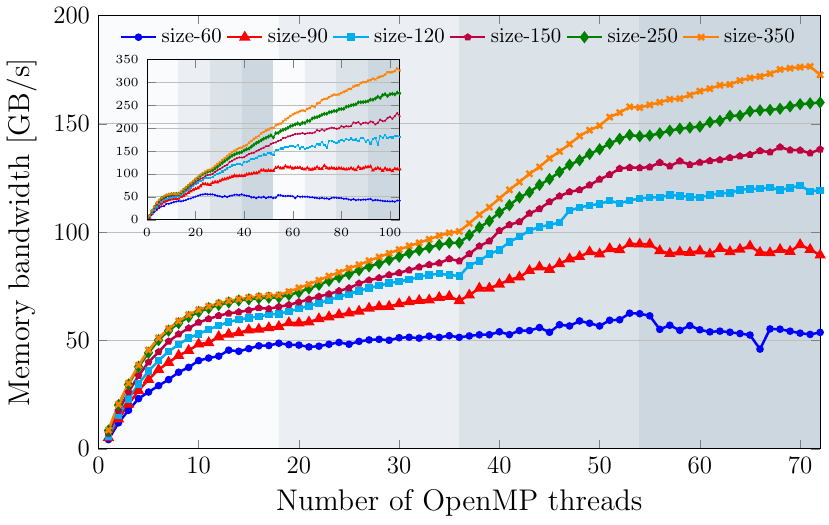}}\quad 
        \subfloat[ Performance in OpenMP] {\includegraphics[scale=0.4]{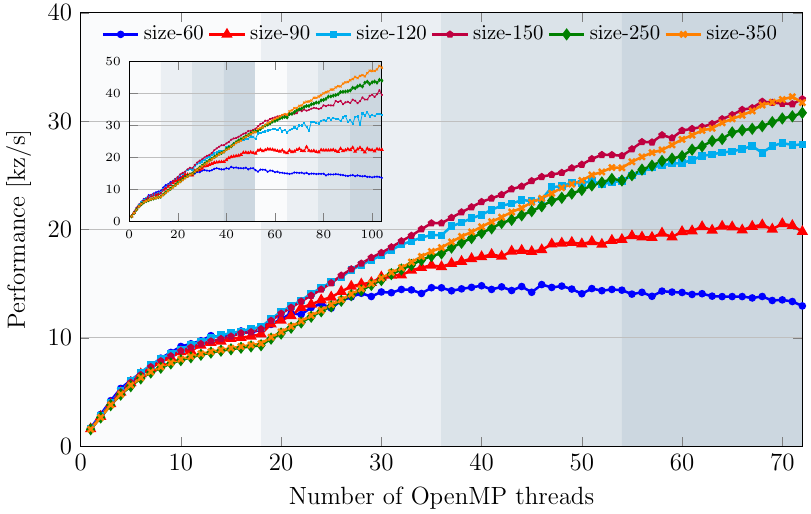}}\quad
        \subfloat[ Overall instructions in OpenMP] {\includegraphics[scale=0.4]{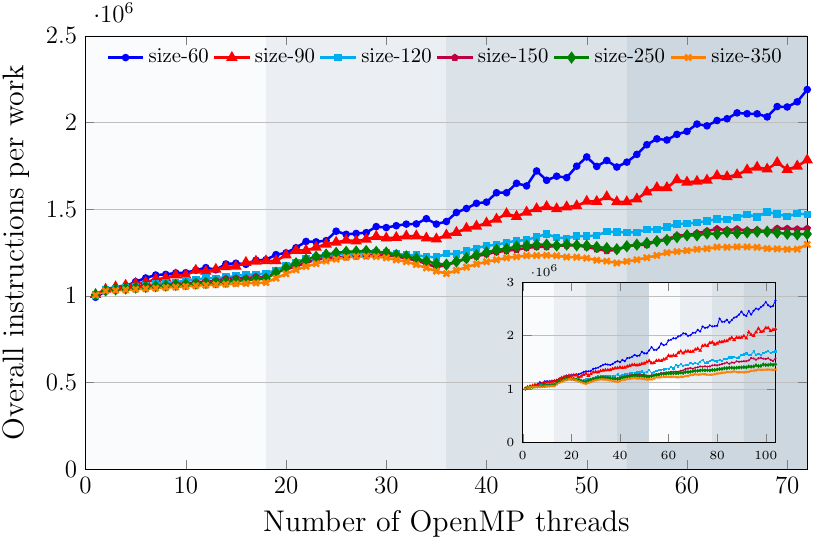}}
    \end{minipage}%
    
    \begin{minipage}{\textwidth}
        \centering
        \subfloat[ Bandwidth in MPI]{\includegraphics[scale=0.4]{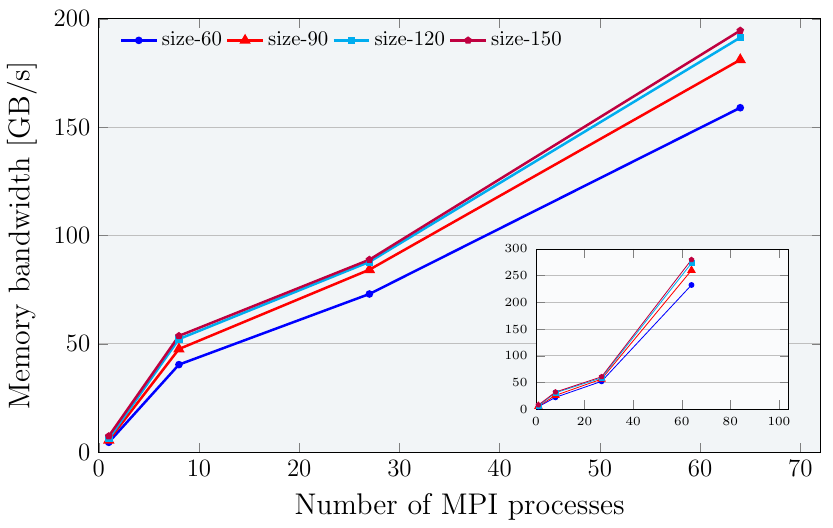}}\quad
        \subfloat[ Performance in MPI]{\includegraphics[scale=0.4]{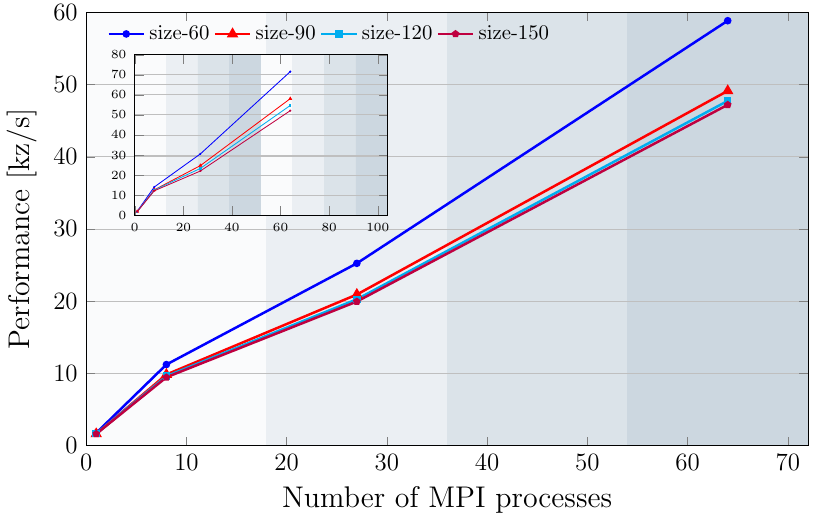}}\quad
        \subfloat[ Overall instructions in MPI]{\includegraphics[scale=0.4]{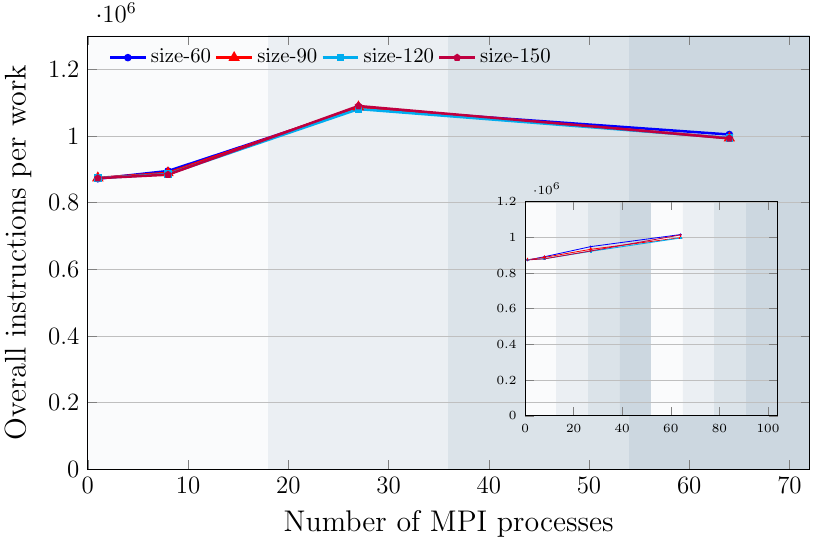}}
    \end{minipage}
    \caption{Performance for the full LULESH application, presented for both OpenMP (top) and MPI (bottom) versions on the ICL-based ClusterA node (subplot view in each plot: SPR-based ClusterB).
    The plots illustrate: (a) memory bandwidth, (b) performance, and (c) the total number of overall instructions normalized by domain size.}
    \label{fig:fullAppAnalysis}
\end{figure*}
\section{Full application analysis}\label{sec:fullapplication}
This section evaluates the performance, power, and energy to solution for the full LULESH application on SPR and ICL. The theoretical ratios of peak performance, memory bandwidth, and thermal design power (TDP) for the Sapphire Rapids node compared to the Ice Lake node on the ClusterA are 1.2, 1.5, and 1.4, respectively.
\subsection{Domain size impact}
On modern architectures, the execution and data transfer features render the LULESH code memory-bound.
In Figure \ref{fig:fullAppAnalysis}(a), memory bandwidth increases significantly up to a domain size of 150, after which the increase becomes less pronounced, reaching \{49, 56, 62, 66, 70, 71\} \GB/s (ICL) and \{41, 47, 51, 54, 57, 57\} \GB/s (SPR) for domain sizes of \{60, 90, 120, 150, 250, 350\}\footnote{
The achievable memory bandwidth for a ccNUMA domain on the Ice Lake system ranges from 75 to 85\,\GBS\ (72\%-82\% of the 104.2\,\GBS\ theoretical bandwidth), while on the Sapphire Rapids system, it ranges from 58 to 62\,\GBS\ (75\%-81\% of the 76.8\,\GBS\ theoretical bandwidth).}.

In Figures \ref{fig:fullAppAnalysis}(a, b), while memory bandwidth saturates by staying roughly constant towards the end of the ccNUMA domain, performance continues to exhibit a slope, suggesting that write-allocate evasion helps reduce traffic.
The OpenMP parallelization of LULESH shows at least partial NUMA-awareness, reaching saturation in the first domain and improving performance in the second domain.
The code intensity (in~\FB) decreases with larger domain sizes.
With a larger domain size of 350, increasing from one to eight ccNUMA domains on SPR's node yields a 6-fold improvement in both memory bandwidth and performance (see subplot inserts), while increasing from one to four ccNUMA domains on ICL's node results in a 3-fold improvement, indicating scalability without additional traffic.
The memory bandwidth on the SPR node of ClusterB (326 GB/s) is 1.9 times that of the Ice Lake node on Cluster A, representing about half (53\%) of the SPR node's theoretical bandwidth and its performance is 1.5 times that of the Ice Lake node on ClusterA. 

The reduced scalability of smaller domain sizes is mainly driven by the OpenMP barrier overhead, where execution stalls at barriers, as shown in Figure \ref{fig:fullAppAnalysis}(c).
To eliminate the straightforward effect of larger problem sizes generating more instructions, we normalized the total instruction count by problem size in OpenMP (Figure \ref{fig:fullAppAnalysis}(c)) and by both problem size and MPI process count in MPI (Figure \ref{fig:fullAppAnalysis}(f)), to further remove weak scaling impact, giving the average instructions executed per process. 
Figure \ref{fig:fullAppAnalysis}(c) shows that the overall average number of instructions executed across all threads rises as scaling progresses from 1 to 72 cores due to OpenMP runtime instructions.
However, the arithmetic instructions remain constant as the workload size remains unchanged.
Crossing the NUMA domain boundary introduces instruction ``bumps,'' indicating that threads in the second NUMA domain must wait at the OpenMP barrier until saturation is achieved.
As OpenMP overhead increases, so does the instructioncount;, theinstructions differences among domainsizes -- initially the same for one OpenMP thread -- expand, making this effect more pronounced for smaller domain sizes.
For instance, when scaling from 1 to 72 threads, the instruction count per data point increases by a factor of 2.1 for the $60^3$ domain size.
This more than doubling the instruction workload, demonstrating the impact of OpenMP overhead as load imbalance was disabled. 
For the largest domain size, $350^3$, the increase factor is 1.3, and \enquote{bumps} could not be avoided due to memory-bound components, meaning that the partial population of additional NUMA domains still results in waiting times.

\begin{figure}[t]
    \begin{minipage}{\textwidth}
        \subfloat[ Power in OpenMP]{\includegraphics[scale=0.3]{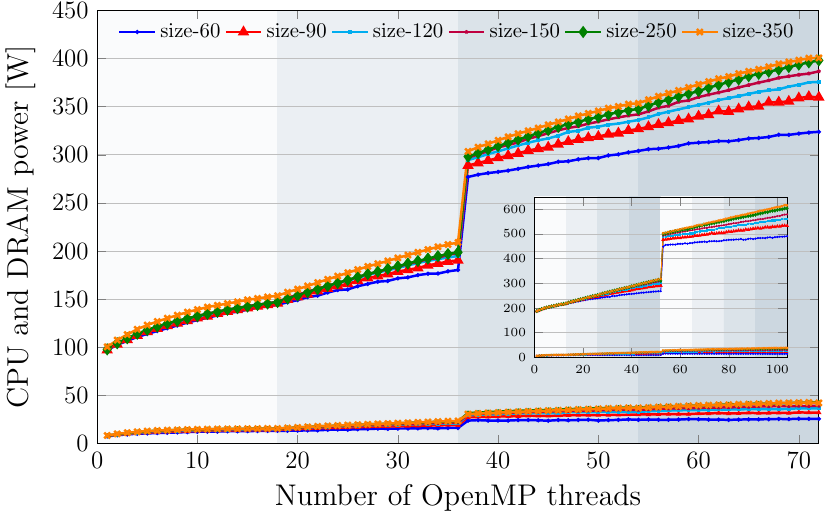}}\quad
        \subfloat[ Energy in OpenMP]{\includegraphics[scale=0.3]{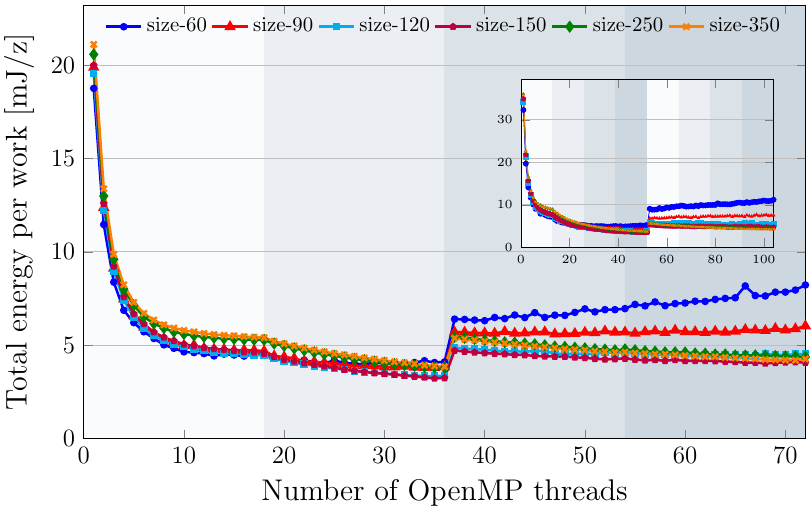}}
    \end{minipage}%
    
    \begin{minipage}{\textwidth}
          \subfloat[ Power in MPI]{\includegraphics[scale=0.3]{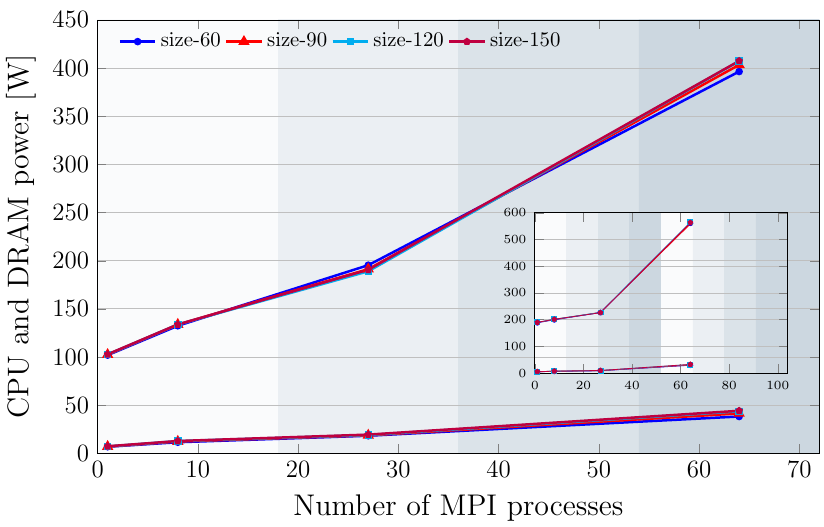}}\quad
        \subfloat[ Energy in MPI]{\includegraphics[scale=0.3]{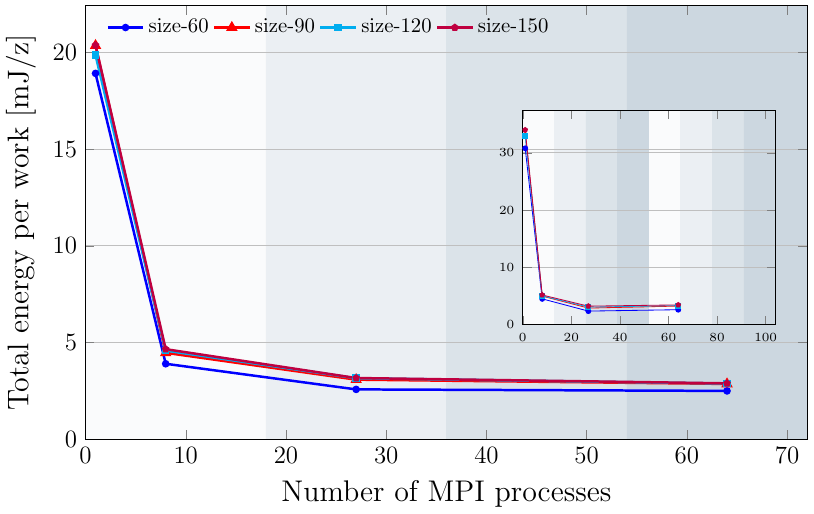}}
    \end{minipage}
    \caption{Performance-energy trade-offs for the full LULESH application, presented for both OpenMP (top) and MPI (bottom) versions on the ICL-based ClusterA node (subplot view in each plot: SPR-based ClusterB).
    The plots illustrate: (a, c) CPU power (top) and DRAM power (down) and (b, d) total energy.}
    \label{fig:fullAppAnalysis-energy}
\end{figure}

In the bottom row of Figure \ref{fig:fullAppAnalysis}, when compared to OpenMP (Figure \ref{fig:fullAppAnalysis}(b)), the MPI implementation demonstrates superior performance on both ICL and SPR platforms (Figure \ref{fig:fullAppAnalysis}(e)).
However, direct comparisons between strong-scaling OpenMP and weak-scaling MPI are challenging due to problem-dependent behavior.
The MPI performance results do not reflect the usual OpenMP pattern where performance improves as the domain size grows.
OpenMP is highly synchronized and incurs substantial overhead due to OpenMP barriers.
Increasing the problem size significantly reduces the OpenMP barrier overhead and also decreases MPI communication overhead.
Ultimately, with a very large problem size, the focus shifts to comparing the implementations of the underlying algorithms.
This allows to discern the influence of the programming model from the influence of the implementation of the code.
Since the performance converges for very large problem sizes, findings indicate that the underlying implementations are effectively equivalent, and the observed differences at smaller problem sizes are due to different overheads.

Increasing the number of MPI processors from 27 to 64 for a $150^3$ domain results in a roughly two-fold rise in performance (2.37 factor) and in total power and energy (2.2 factor) on both ICL and SPR nodes; see Figures \ref{fig:fullAppAnalysis}(e) and \ref{fig:fullAppAnalysis-energy}(c, d).
However, the memory bandwidth 
on SPR rather increases by 4.6 times than twice as on ICL; see Figure \ref{fig:fullAppAnalysis}(d).
This is due to the problem size increase under weak scaling, which remains small enough for up to 27 processes to fit into the larger aggregate cache of Sapphire Rapids compared to Ice Lake. As a result, the MPI-parallelized implementation of LULESH is not entirely memory-bound on Sapphire Rapids. 
Due to the cache effect outweighing communication overhead in the smaller $60^3$ domain size (blue) that fits into the cache, lower memory bandwidth and improved performance lead to minimal energy to solution and EDP.

In Figure \ref{fig:fullAppAnalysis-energy}(a), on-chip and DRAM power sharply increase at the socket changeover (36 cores) due to added baseline power (roughly 90 W for ICL and 180 W for SPR).
This socket switchover impact from compact pinning is trivial, and spreading the pinning across both sockets eliminates it.
As expected, overall energy decreases while power increases with more cores.
LULESH is a ``cold'' application, consuming on-chip power of 80\% and 86\% of TDP for a full SPR and Ice Lake node, respectively. 
However, the DRAM contributes only approximately 6\% (SPR) to 10\% (ICL) to the total energy or power consumption \cite{AfzalHW:2023:2}.
This is because, compared to the DDR4 on Ice Lake, the DDR5 on SPR is far less power-hungry and has a considerably smaller overall impact.
Simple on-chip and DRAM power model\footnote{On-chip and DRAM power increase linearly with active cores until a bottleneck occurs. After that, inactive cores wait for memory, causing on-chip power to rise slowly without improving performance, while DRAM becomes constant.} hold and SPR node consumes 48\% more total power than Ice Lake system.

In Figure \ref{fig:fullAppAnalysis-energy}(b), beyond a single socket, as scaling declines sharply, the energy consumption on the second socket stays constant -- showing no further decrease -- with a noticeable shift at the switchover point.
Given SPR's higher peak performance and greater memory bandwidth compared to ICL, its performance scales proportionally, falling somewhere within the ratio of peak performance and memory bandwidth.
For memory-bound codes, the theoretical increase of 40\% (44.5\% measured) in power consumption is offset by the 50\% increase in memory bandwidth, resulting in comparable energy-to-solution on both systems but yielding lower EDP values on the SPR ccNUMA domain than Ice Lake.
For example, energy consumption for the OpenMP implementation is 4.2 mJ/z on both the SPR node and the ICL node; see Figure \ref{fig:fullAppAnalysis-energy}(b). 

\smallskip \highlight{\emph{Upshot}: 
To isolate the impact of programming model choices from code implementation details, a comparison across different problem sizes shows that OpenMP overheads are more significant for smaller problem sizes.
The general characteristics of the code are similar on both ICL and SPR architectures, while detailed performance and energy comparisons are specific to these chips.}

\begin{figure}[t]
    \begin{minipage}{0.95\textwidth}
    \hspace{-7pt}
        \subfloat[ Performance in MPI]{\includegraphics[scale=0.3]{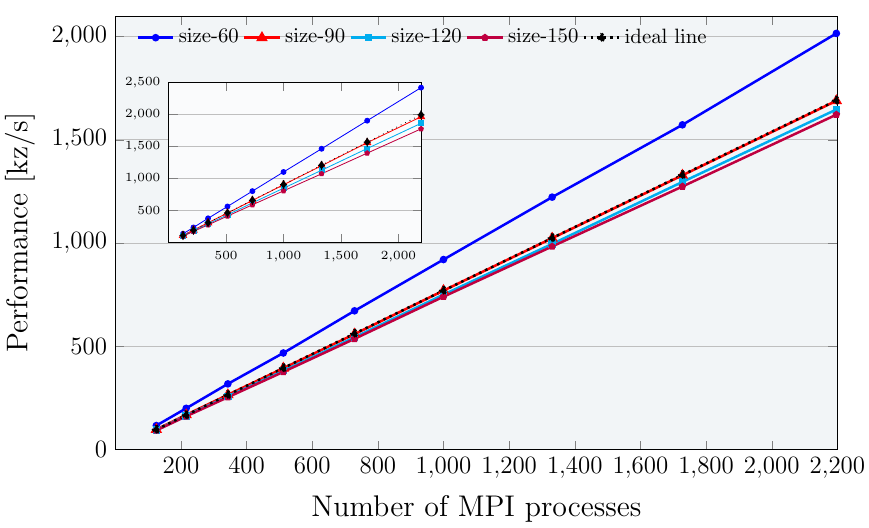}}\quad
        \subfloat[ Bandwidth in MPI]{\includegraphics[scale=0.3]{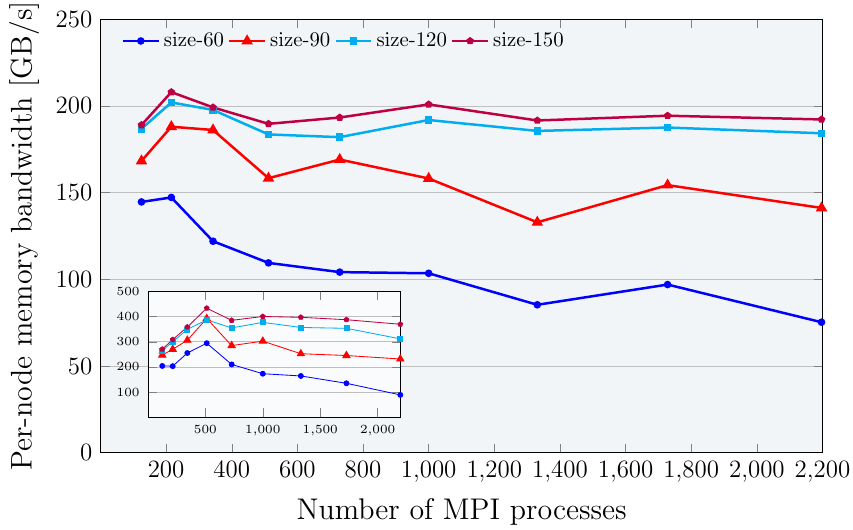}}     
    \end{minipage}%
    
    \begin{minipage}{0.95\textwidth}
          \subfloat[ Power in MPI]{\includegraphics[scale=0.3]{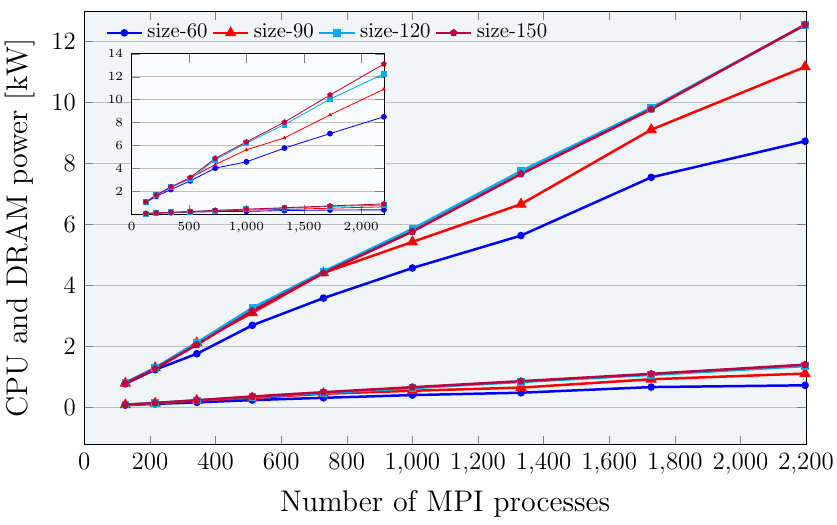}}\quad
        \subfloat[ Energy z-plot in MPI]{\includegraphics[scale=0.3]{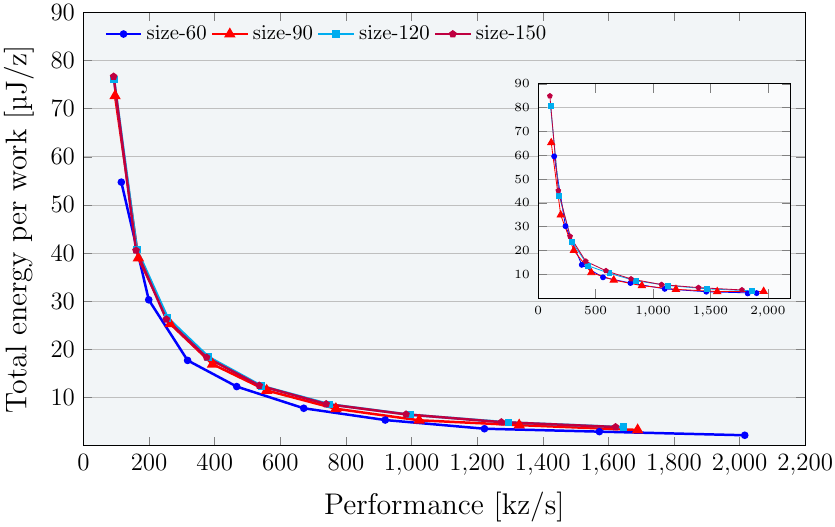}}
    \end{minipage}
    \caption{Weak scaling multi-node runs for the MPI-parallel LULESH application on the ICL-based ClusterA (subplot view in each plot: SPR-based ClusterB).
    The plots illustrate: (a) performance, (b) bandwidth, and (c) CPU power (top) and DRAM power (down) (d) energy, when scaling till 2197 processes (31 nodes of ClusterA and 22 nodes of ClusterB).}
    \label{fig:fullAppAnalysis-cluster}
    \vspace{1em}
\end{figure}

\subsection{Multi-node weak scaling impact}
Figure \ref{fig:fullAppAnalysis-cluster}(a) shows near-perfect weak scaling beyond a single node when scaling up to 31 ClusterA nodes or 22 ClusterB nodes ($5^3$ to $13^3$ processes). 
The actual speedup closely aligns with the ideal speedup (based on a $5^3$ MPI processor baseline), classifying LULESH as a scalable code.
Within a node, memory bandwidth grows and the slight reduction in scaling is primarily caused by bandwidth sharing, as noted earlier.
Figure \ref{fig:fullAppAnalysis-cluster}(b) shows that with higher node counts, the per-node memory bandwidth remains nearly consistent for larger domains but gradually decreases for smaller domains.
This deviation from the ideal horizontal line represents a loss in parallel efficiency.
For example, for a $90^3$ domain size, bandwidth drops to 22\% (ICL) and 11\% (SPR).

In Figure \ref{fig:fullAppAnalysis-cluster}(c), on-chip power dissipation reaches 12.6 kW (81\% of 15.5 kW TDP) for 31 ClusterA nodes and 13.1 kW (85\% of 15.4 kW TDP) for 22 ClusterB nodes.  
To mitigate the trivial dependency of larger resources consuming more energy in week scaling, energy is normalized by the amount of work done. 
This is illustrated by plotting energy per work unit [J/z] on the y-axis and performance [z/s] on the x-axis while scaling resources for increasingly larger problems, as shown in Figure \ref{fig:fullAppAnalysis-cluster}(d).
The figure shows that good code scalability beyond the node level offsets the high power consumption of 81-85\% of TDP, reducing the overall energy required per work unit.
The inset subplot shows the same energy consumption but lower EDP for SPR-based ClusterB nodes compared to ICL-based ClusterA nodes, as expected.
In contrast, the smaller $60^3$ domain size in blue, which fits into cache, presents an interesting case. It shows a more pronounced per-node memory bandwidth drop, reaching approximately 53\% on ICL and 62\% on SPR.
However, as the cache effect reduces communication overhead, this leads to minimal energy to solution and EDP, resulting in both lower power dissipation and improved performance.

\smallskip \highlight{\emph{Upshot}: 
Beyond a single node, LULESH shows near-perfect scaling and consistent per-node memory bandwidth for larger domains, offsetting high power consumption and yielding reduced energy to solution and EDP.
}

\subsection{Vectorization impact}
Performance was evaluated by progressively enabling vectorization flags (\code{-xSSE4.2}, \code{-xCORE-AVX2}, \code{-xCORE-AVX512}) on the default scalar settings.
The trivial trend that slower, non-vectorized LULESH code scales better than faster, vectorized code was observed.
However, the performance gap was minimal, 
as only ~6\% of the vectorized code utilized AVX instructions. 
The contributions for {scalar, 128-packed, 256-packed, 512-packed} double instructions, as illustrated in Table \ref{tab:vectorization}, remain consistent across domain sizes and thread counts, except under MPI weak scaling.

\input{figures/FullApp/tab_vectorization}

In the OpenMP variant, the sole non-memory-bound, non-vectorized function, \code{calckinematicsForElems}, presents optimization potential.
In contrast, in the MPI variant, four of five non-memory-bound functions lack vectorization and have potential for optimization, with only the \code{UpdateVolumesForElems} non-hot spot function achieving 99.9\% vectorized instructions.
For example, the \code{calckinematicsForElems} hot spot accounts for about 13.2\% (OMP-ICL), 14.3\% (OMP-SPR), 14.9\% (MPI-ICL) and 14.2\% (MPI-SPR) of the overall  execution time, suggesting that enhanced vectorization could boost performance by up to a factor of 4 to 8 (10\% on the ccNUMA level).
Similarly, the MPI hot spot \code{IntegrateStressForElems}, consuming 13.6\% (MPI-ICL) and 12.5\% (MPI-SPR) of execution time, offers similar optimization potential.
In contrast, optimizing non-hot spot functions like \code{CalcCourantConstraintForElems}, \code{CalcHydroConstraintForElems}, and \code{UpdateVolumesForElems}, which contribute only 0.8\%, 0.3\%, and 0.2\% of total execution time, respectively, is unlikely to provide meaningful benefits.

\smallskip \highlight{\emph{Upshot}: 
The performance gap between scalar and vectorized code (though minimal with only 6\% AVX instructions) highlights optimization potential by vectorizing one OpenMP and two MPI non-memory-bound, non-vectorized hot spots.
}

\subsection{Turbo mode impact}
\begin{figure}[t]
    \centering
    \subfloat[ Average clock frequency]{\includegraphics[scale=0.3]{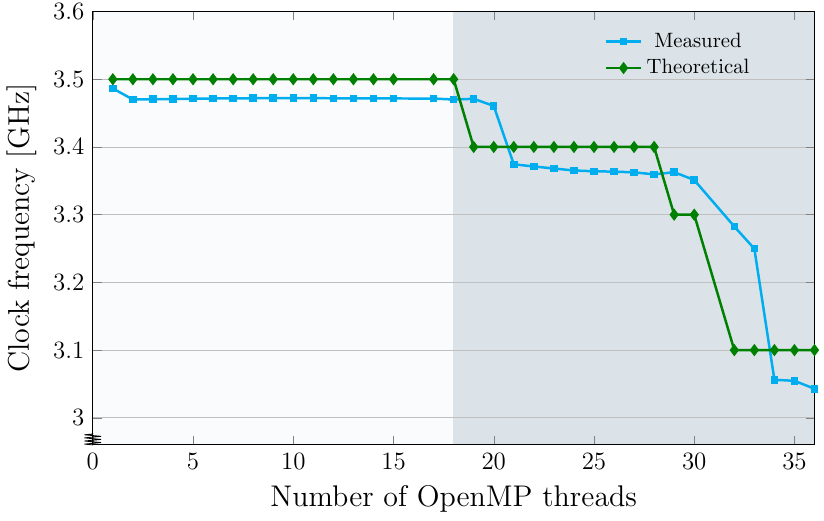}}\quad
    \subfloat[ Performance and energy]{\includegraphics[scale=0.3]{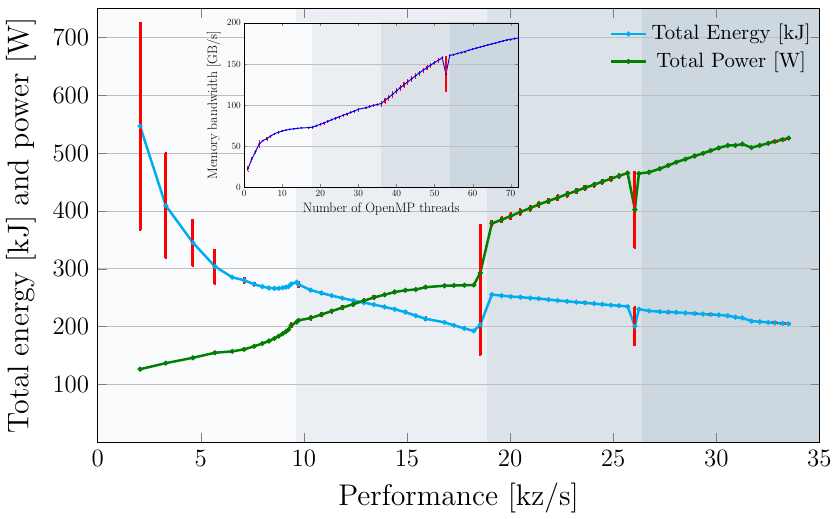}}
    \caption{Influence of turbo mode on memory bandwidth, performance, power and energy for the OpenMP-parallel LULESH application on ClusterA' node. 
    (a) The 3.48 \GHZ~frequency of first domain decreases in the second domain.
    (b) The minimum, maximum and average statistics from ten turbo mode runs are shown.}
    \label{fig:fullAppAnalysis-turbo}
    \vspace{15pt}
\end{figure}

In Figure \ref{fig:fullAppAnalysis-turbo}(a), results focus on the first chip, as data beyond it provides no new insights.
On the second chip, the average frequency for all cores rises until saturation (69–72 cores), after which they stabilize at 3.08 \GHZ, dropping from 3.19 \GHZ.
Measurements align with theoretical turbo tables of ICL and SPR obtained using \code{likwid-powermeter}.
In SPR turbo mode, a similar trend is observed: 3.8 \GHZ~in the first domain, dropping from 3.6 \GHZ~to 3.0 \GHZ~in the second. For brevity, only ICL results are shown, as SPR results do not differ qualitatively.

In Figure \ref{fig:fullAppAnalysis-turbo}(b), warm-up steps were incorporated, and readings were repeated over 10 iterations, showing the maximum (top horizontal bars), minimum (bottom horizontal bars), and average (data points) statistics.
An Ice Lake core in turbo mode reaches up to 3.5 \GHZ~(46\% above base frequency), and a node reaches 3.07 \GHZ~(28\% above base frequency), with a frequency boost occurring at chip switchover.
Consequently, compared to the base frequency, turbo mode increases performance, memory bandwidth, on-chip power, and on-chip energy by approximately 6.7\%, 7\%, 22.8\% (91.2\% of the TDP), and 14.36\%, respectively. 
\smallskip \highlight{\emph{Upshot}: 
Based on the Energy-Delay Product metric, turbo mode may not be the optimal choice, as the energy cost increases by approximately twice the performance gain.
}
\begin{figure*}[t]
    \centering
    \subfloat[ Power in OpenMP]{\includegraphics[scale=0.4]{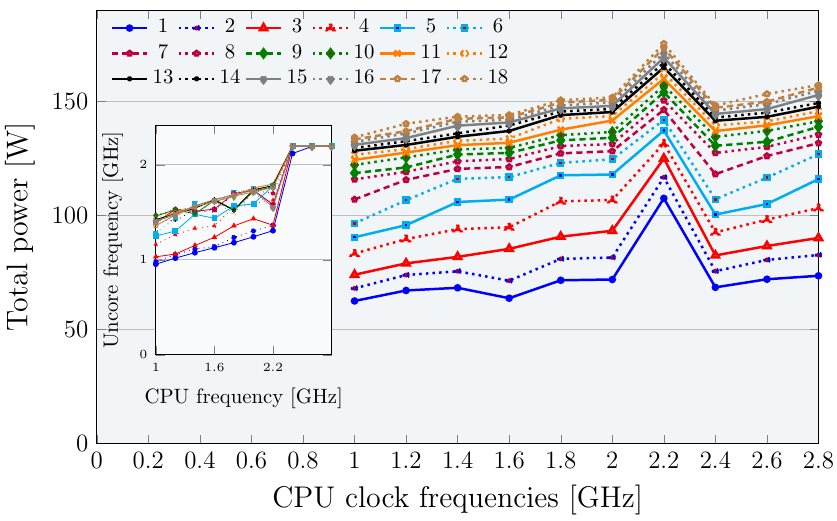}}\quad
    \subfloat[ Energy z-plot in OpenMP]{\includegraphics[scale=0.4]{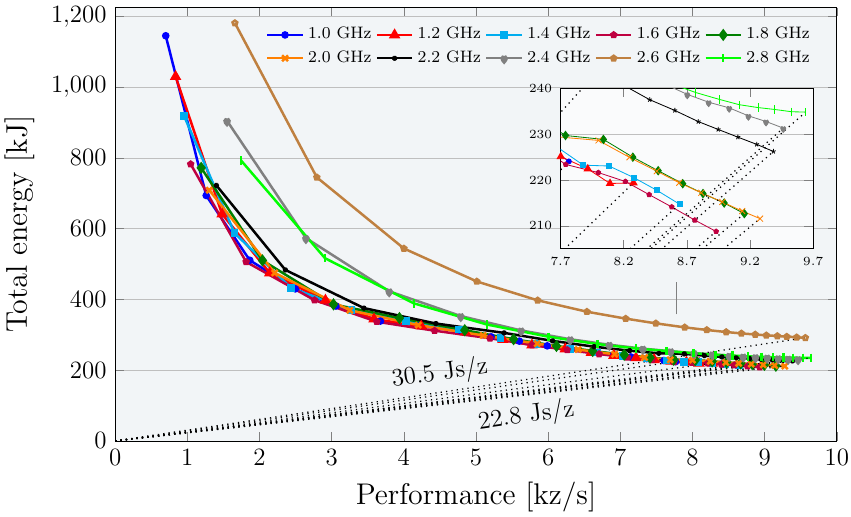}}\quad
    \subfloat[ Energy z-plot in MPI]{\includegraphics[scale=0.4]{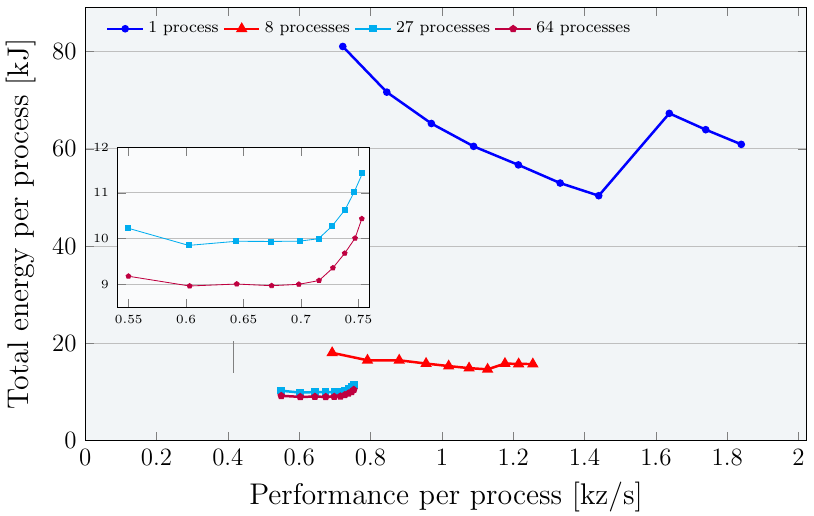}}
    \caption{
    The influence of CPU clock frequency on the full LULESH application is presented for both (a, b) OMP-ICL and (c) MPI-ICL variants. The second and third plots present a z-plot of total energy-to-solution versus performance, highlighting the optimal Energy Delay Product for various cores and core frequencies (1.0-2.8 \GHZ). 
    The uncore frequency remains unfixed, ranging from a minimum of 0.8 \GHZ~to a maximum of 2.5 \GHZ.
    The subplots in (b-c) display zoomed-in insets, while the subplot in (a) illustrates how the uncore frequency adjusts internally when the CPU frequency is fixed.}
    \label{fig:fullAppAnalysis-zplot}
\end{figure*}
\subsection{Core frequency impact}
Figure \ref{fig:fullAppAnalysis-zplot}(a) suggests a linear relationship between total power $W$ and CPU frequency $f$, expressed as
\bq
W = W_0 + W_d f\cma
\eq
where the zero-frequency baseline power $W_0$ ranges from 40-120 W depending on the number of active cores, and $W_d$ represents dynamic power.
In contrast, the zero-core baseline power ranges from 55-95 W depending on the CPU clock frequency, out of which baseline DRAM power is limited to a narrow range of 3-5 W.
These baseline (idle) powers were determined through linear regression for the CPU and curve fitting for DRAM, accounting for approximately 16\%-48\% of socket' TDP and 19\%-57\% of socket' total power.
With the BIOS uncore frequency range set to 0.8–2.5 \GHZ~and the performance-energy bias at 15 (lowest energy), uncore frequency adjusts dynamically when not explicitly set, as shown in the subplot inset of Figure \ref{fig:fullAppAnalysis-zplot}(a).
This causes a significant increase in power and energy consumption at a core frequency of 2.2 \GHZ, especially at lower core counts, as shown in Figure \ref{fig:fullAppAnalysis-zplot}(a)-(c).
Since SPR findings are qualitatively similar and do not provide additional insights, we only present the ICL results for both OpenMP (Figures \ref{fig:fullAppAnalysis-zplot}(a)-(b)) and MPI variants (Figure \ref{fig:fullAppAnalysis-zplot}(c)) for brevity.

In Figure \ref{fig:fullAppAnalysis-zplot}(b), energy results for CPU frequencies below the base frequency are closely aligned, while CPU frequencies above the base frequency lead to higher energy consumption.
For a single ccNUMA domain, the lowest CPU energy consumption occurs at 1.6 \GHZ, whereas 2.0 \GHZ~achieves optimal energy efficiency (minimum EDP).
This is because uncore frequencies are lower below 2.2 \GHZ~and higher above it.
If one is able to pay an additional energy, operating at 2.0 \GHZ~frequency is optimal.
However, if a power cap is required, this will be different, as lower CPU frequencies burn less power than 2.0 \GHZ.

In Figure \ref{fig:fullAppAnalysis-zplot}(c), energy and performance are normalized by process count to isolate trivial weak-scaling effects, where more processes increase performance and energy use due to larger problem sizes.
Similar to OpenMP results, exceeding a 2.0 \GHZ~CPU frequency wastes energy with marginal performance gains, raising EDP. 
For example, increasing frequency from 2.0 \GHZ~to 2.8 \GHZ~with 64 processes (purple) yields only 5\% more performance at a 15\% energy cost.
\smallskip \highlight{\emph{Upshot}: 
Adjusting the CPU clock frequency causes a 23\% (5\%) variation in performance and a 40\% (15\%) variation in energy consumption for OpenMP (MPI), highlighting the significant effect of core frequency on energy with 2.0 \GHZ~identified as optimal. Though the impact remains relatively limited for lower frequencies.
}

\subsection{Uncore frequency impact}
\begin{figure*}[t]
    \centering
    \subfloat[ Performance]{\includegraphics[scale=0.4]{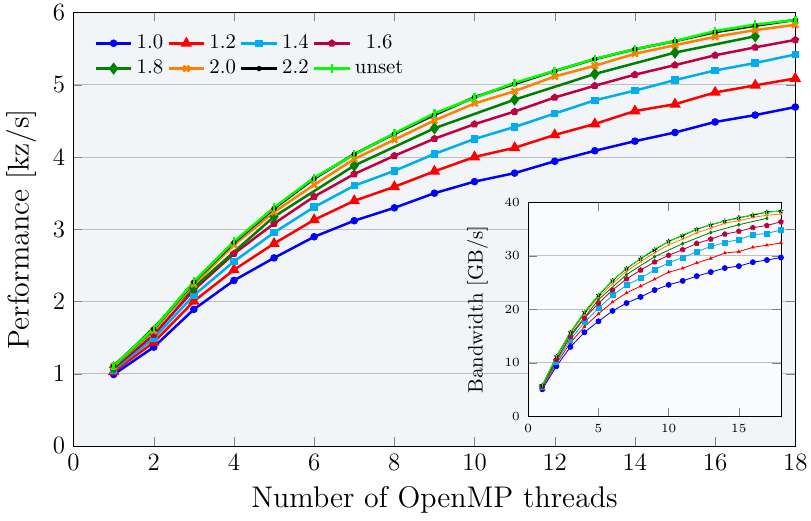}}\quad
    \subfloat[ Power]{\includegraphics[scale=0.4]{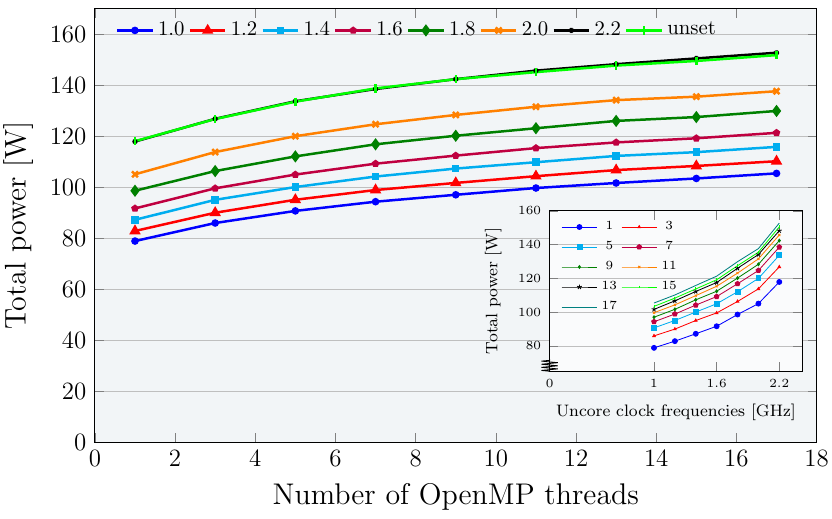}}\quad
    \subfloat[ Energy z-plot]{\includegraphics[scale=0.4]{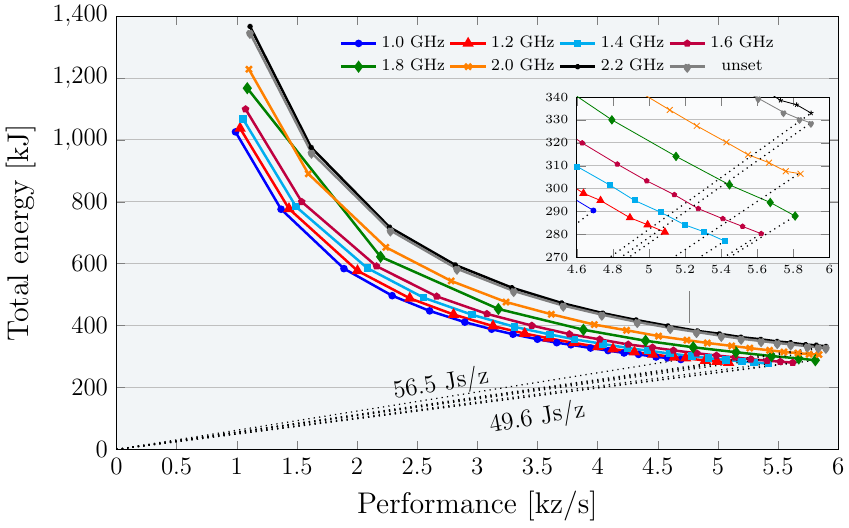}}
    \caption{Influence of uncore clock frequency on (a) performance (subplot view: bandwidth), (b) power vs. cores (subplot view: power vs. uncore frequency) and (c) energy z-plot (subplot view: zoom-in) for the OpenMP-parallel LULESH application on a ccNUMA domain of an ICL domain. 
    The core clock frequency is set to the base frequency of 2.4 \GHZ, while the uncore clock frequency is either left unset (min/max 2.4 \GHZ) or allowed to fix between 1.0--2.2 \GHZ.}
    \label{fig:fullAppAnalysis-ICX36}
\end{figure*}

Figure \ref{fig:fullAppAnalysis-ICX36} presents results with uncore frequencies either left unset or fixed between 1.0 \GHZ~and 2.2 \GHZ~within a ccNUMA domain.
As uncore frequency fixing is unavailable on ClusterA, the icx36 node from another test cluster was used, with uncore frequency configured via the \code{likwid-setFrequencies} tool. 
For brevity, we focus on the OMP-ICL variant, as other findings are similar.
Figure \ref{fig:fullAppAnalysis-ICX36}(a, b) shows that downclocking the uncore frequency reduces memory bandwidth by 29\%, performance by 26\%, and saves 42\% power. 
This drop in memory bandwidth suggests it is not solely due to the lowered uncore frequency; rather, the LULESH code slows down as it comprises non-memory-bound functions (relying on L3 cache bandwidth influenced by uncore frequency) in addition to memory-bound functions.
The performance reduction observed on the ICX36 node compared to the ClusterA node indicates that the uncore frequency, when left unset on ClusterA, operates at a significantly higher value, which enhances L3 cache speed and boosts performance.
The ICX36 node operates slower due to its BIOS configuration, including an uncore frequency range of 2.4–2.4 \GHZ~(vs. 0.8–2.5 \GHZ) and a performance-energy bias of 6 (middle of performance/energy scale) instead of 15 (lowest energy).

Figure \ref{fig:fullAppAnalysis-ICX36}(b) shows that the baseline power, ranging from 75–110 W (zero cores) to 65–90 W (zero frequency) across uncore clock frequencies with a fixed base CPU frequency on a ccNUMA domain, is more accurately represented by the zero-core value.
The baseline (idle) power, accounting for 26–44\% of the socket's TDP, was estimated through linear regression by extrapolating to zero core or zero frequency.
The subplot inset (for different core counts) shows that the linear frequency-power relationship holds up to 1.6 \GHZ.
The observed non-linearity beyond this point is likely due to the CPU dynamically adjusting voltage with frequency to ensure proper chip operation, possibly as a preset for consistent regulation.
At lower frequencies, the voltage decreases, but when the frequency drops to a level where voltage can no longer be reduced further, it remains constant, resulting in a linear power-frequency relationship. However, at higher frequencies above 1.6 \GHZ, voltage increases, leading to a non-linear trend.
The noticeable increase in energy and power observed at CPU frequencies above 2.2 \GHZ, when the uncore frequency is unfixed on ClusterA (Figure \ref{fig:fullAppAnalysis-zplot}) is an influence of varying uncore frequency, and thus vanishes in on-chip measurements with a fixed uncore frequency; however, this does not hold for DRAM measurements. 

Figure \ref{fig:fullAppAnalysis-ICX36}(c) shows the global minimum energy-to-solution occurs at a uncore frequency of 1.4 \GHZ~for the full ccNUMA domain (1 \GHZ~for the single core), with only a 19\% (34\%) variation between the minimum and maximum.
These variations from reducing the uncore frequency are lower than those in the previous Intel generation \cite{Hofmann:2018}.
As a result, the entire domain (chip in general) should be used to minimize energy. 
Notably, in the highest 2.2 \GHZ~uncore frequency scenario, a high baseline power drives the Energy-Delay Product to its peak.
Increasing the uncore frequency boosts memory bandwidth until it saturates around 2.0 \GHZ.
In the 1.6-1.8 \GHZ~range of uncore frequency, there is still some leeway where memory bandwidth decreases less with decreasing uncore frequency, offering power savings and a minimum EDP.
\smallskip \highlight{\emph{Upshot}: 
Adjusting the uncore clock frequency can result in a 26\% variation in performance and a 20\% variation in energy consumption, suggesting that fixing the uncore frequency has minimal impact, with 1.6-1.8 \GHZ~being advisable.}
\section{Summary and future work} \label{sec:conclusion}
We presented an in-depth performance and energy analysis of the LULESH proxy application on two multi-core clusters with Intel Ice Lake (ICL) and Sapphire Rapids (SPR) processors, comparing OpenMP and MPI implementations.
Each analysis section, both at the full application level and for individual kernels level, provides insights that collectively reveal LULESH's performance and energy limitations, architectural impacts, and optimization potential.

\paragraph{Key takeaways}
In the kernel analysis, LULESH consists of six hot spot functions, whose runtime contribution is similar between the OpenMP and MPI implementations. However, four memory-bound functions in OpenMP exhibit scalable behavior in the MPI version, including one hot spot function, making them strong candidates for in-core optimizations rather than for memory data traffic reduction strategies. Compared to the full LULESH application within a single ccNUMA domain, all six hot spot functions converge within the same power range, consuming up to 58\% of the total energy. 
Using the Roof{}line model, we provided performance limits for each memory-bound hot spot function. Using hardware performance counter events, we showed that actual measurements fall within the range between the best-case and worst-case predictions (achieved with write-allocate evasion and data reuse).

In the full application analysis for both ICL and SPR architectures, the code exhibits partly memory-bound characteristics, while the detailed comparison of energy and performance is specific to these architectures.
To separate the impact of programming model choices from code implementation specifics, we compare the performance of OpenMP and MPI implementations across various problem sizes, noting that OpenMP implementation overheads become particularly significant for smaller problem sizes.

Hypothetical idle power consumption, defined as the extrapolated power consumption at zero active cores or zero frequency, is a critical factor for both CPUs compared to older Intel designs and is especially high on SPR, making concurrency throttling (using fewer cores for memory-bound code) less effective.
A maximum of 40\% variation in energy consumption was observed with core frequency adjustments and 20\% with uncore frequency adjustments.
Therefore, energy to solution can be reduced for LULESH by varying the number of cores, the core clock frequency, and the uncore clock frequency, although the impact of all of them is limited.
For the performance-energy trade-off, LULESH exhibits a lower EDP on SPR than on ICL due to a complex interplay among SPR's superior memory bandwidth, higher core count, higher power dissipation, and lower clock speed.
While our results reflect the current performance and energy trade-offs on Intel CPUs, the metrics, analytical methods and insights employed are broadly applicable in similar high-performance applications.

\paragraph{Future work}
In future work, we aim to expand this analysis to hybrid parallelization strategies beyond pure MPI and pure OpenMP in order to further provide comprehensive insights into performance-energy trade-offs and optimization opportunities in LULESH and similar applications.
Additionally, we anticipate interesting results from studying idle wave~\cite{AfzalHW:2019,AfzalHW:2021} and desynchronization~\cite{AfzalHW:2020,AfzalHW:2022:1,AfzalHW:2022:2,AfzalHW:2022:4,AfzalHW:2023:1} phenomena in LULESH, which exhibits a mix of memory- and compute-bound behavior.
We will assess the performance of LULESH on architectures equipped with HBM or integrated accelerators or AI capabilities with traditional DRAM setups to clarify the benefits and potential synergies between computational and memory-bound tasks for LULESH and similar memory-intensive applications. 

\ifblind
\else
\section*{Acknowledgments}
This research work is supported by EEC, a central NHR initiative focused on enhancing energy efficiency and managing operational costs across NHR centers.
The authors gratefully acknowledge the 
HPC resources provided by the Erlangen National High Performance Computing Center (NHR@FAU) of the Friedrich-Alexander-Universität Erlangen-Nürnberg. 
NHR funding is provided by federal and Bavarian state authorities. 
NHR@FAU hardware is partially funded by the German Research Foundation (DFG) -- 440719683.
\fi

\bibliographystyle{IEEEtran}
\bibliography{references}

\begin{IEEEbiography}[{\includegraphics[width=1in,height=1.25in,clip,keepaspectratio]{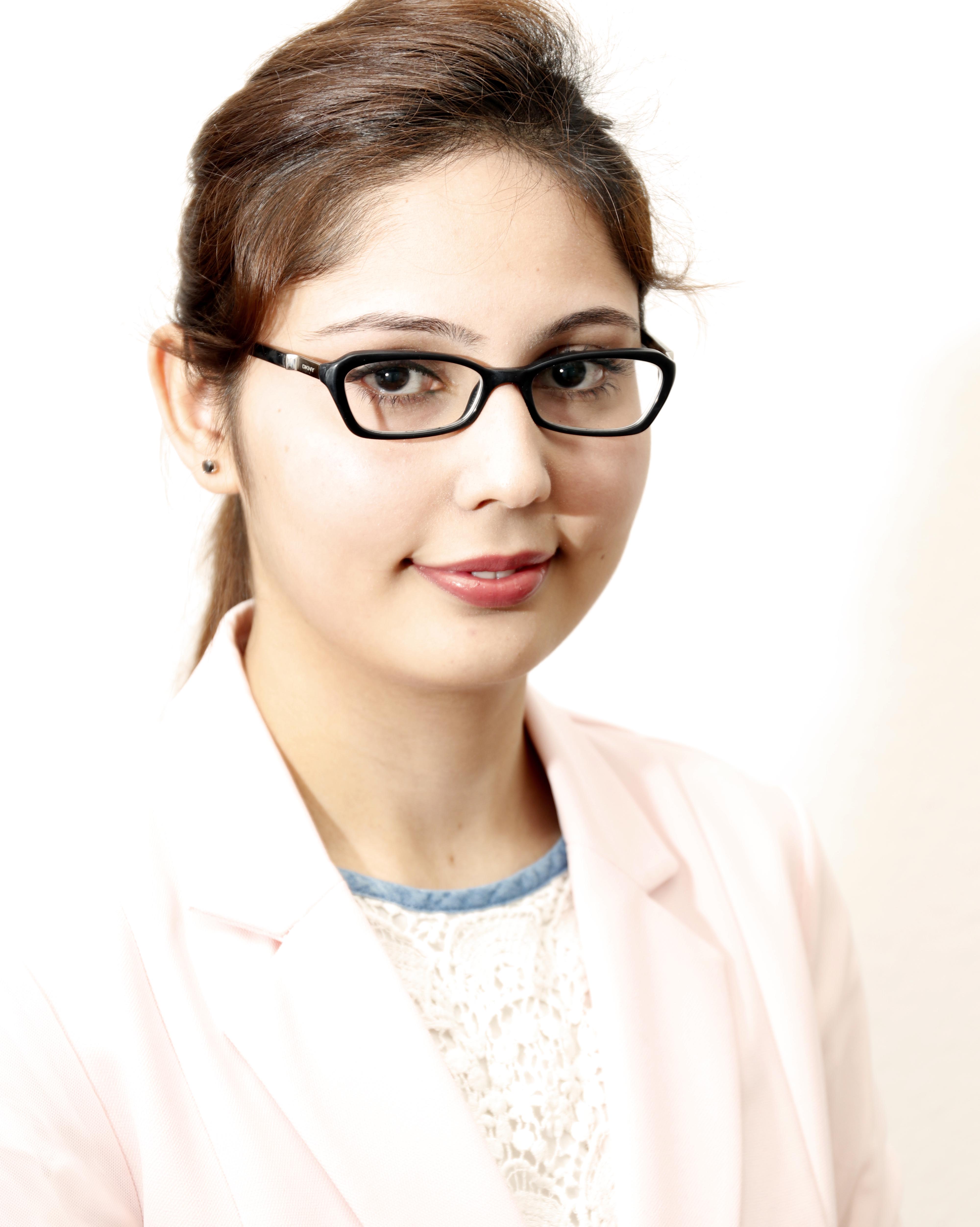}}]{Ayesha Afzal} holds a Master's (M.Sc.) degree in Computational Engineering from Friedrich-Alexander-Universit\"at Erlangen-N\"urnberg, Germany, and a Bachelor's (B.Sc.) degree in Electrical Engineering from the University of Engineering and Technology, Lahore, Pakistan. She is currently completing her Ph.D. at the Erlangen National High Performance Computing Center (NHR@FAU), Germany. Her research interests include analytic first-principles models, analysis tools, and simulation frameworks, with a focus on performance engineering and energy efficiency of distributed-memory parallel programs in high-performance computing.
\end{IEEEbiography}
\begin{IEEEbiography}[{\includegraphics[width=1in,height=1.25in,clip,keepaspectratio]{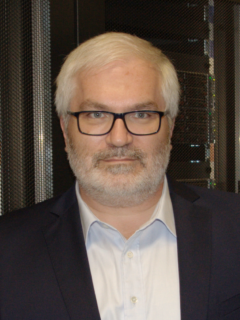}}]{Georg Hager} holds a doctorate (Ph.D.) and a Habilitation degree in Computational Physics from the University of Greifs\-wald, Germany. He leads the Training \& Support Division at Erlangen National High Performance Computing Center (NHR@FAU) and is an associate lecturer at the Institute of Physics at the University of Greifs\-wald. Recent research includes architecture-specific optimization strategies for current microprocessors, performance engineering of scientific codes on chip and system levels, and the modeling of out-of-lockstep behavior in large-scale parallel codes.
\end{IEEEbiography}
\begin{IEEEbiography}[{\includegraphics[width=1in,height=1.25in,clip,keepaspectratio]{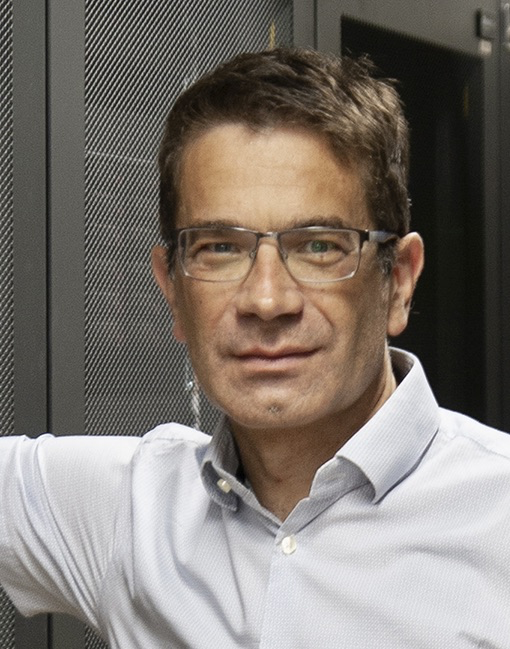}}]{Gerhard Wellein}
 received the Diploma (M.Sc.) degree and a doctorate (Ph.D.) degree in Physics from the University of Bayreuth, Germany. He is a Professor at the Department of Computer Science at Friedrich-Alexander-Universit\"at Erlangen-N\"urnberg and heads the Erlangen National Center for High-Performance Computing (NHR@FAU). His research interests focus on performance modeling and performance engineering, architecture-specific code optimization, and hardware-efficient building blocks for sparse linear algebra and stencil solvers.
\end{IEEEbiography}

\newpage
\appendices
\section{Frequency impact on hot spot functions}\label{App:A}

\begin{figure}[h]
    \begin{minipage}{\textwidth}
        \subfloat[ Perf. (memory-bound)] 
        {\includegraphics[scale=0.3]{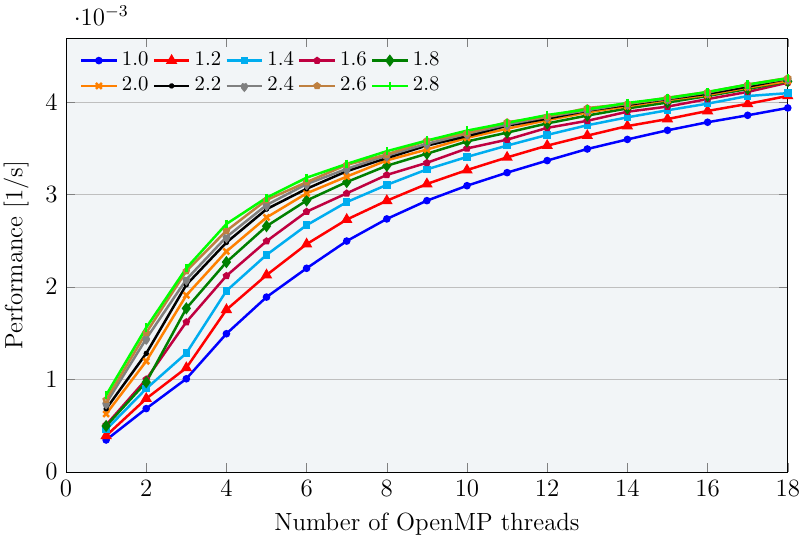}}\quad
        \subfloat[ Perf. (scalable)] 
        {\includegraphics[scale=0.3]{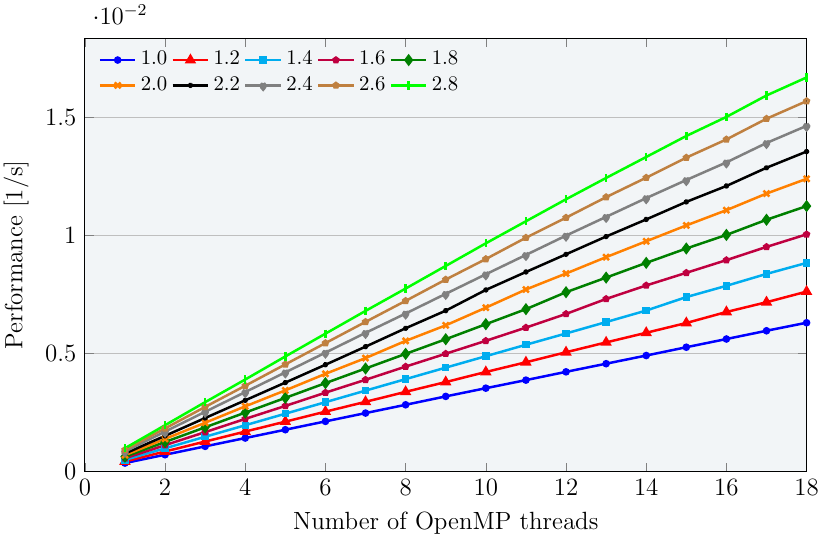}}
    \end{minipage}%
    
    \begin{minipage}{\textwidth}
        \subfloat[ Power (memory-bound)] 
        {\includegraphics[scale=0.3]{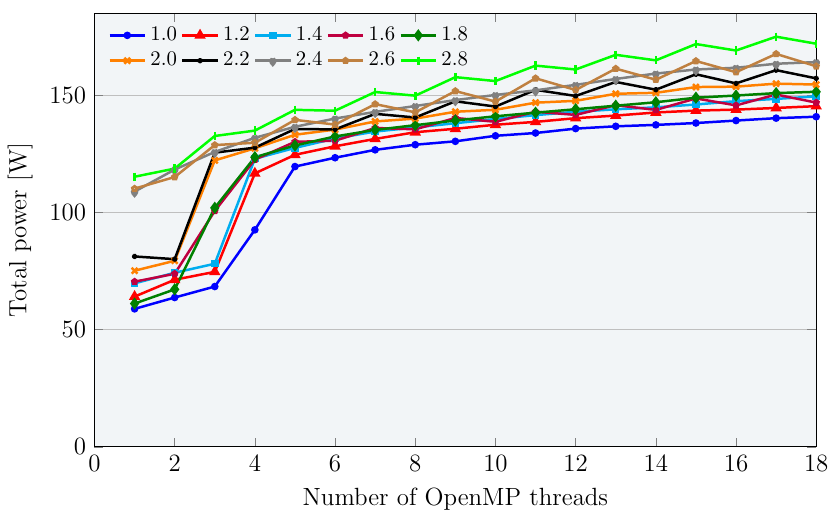}}\quad
        \subfloat[ Power (scalable)] 
        {\includegraphics[scale=0.3]{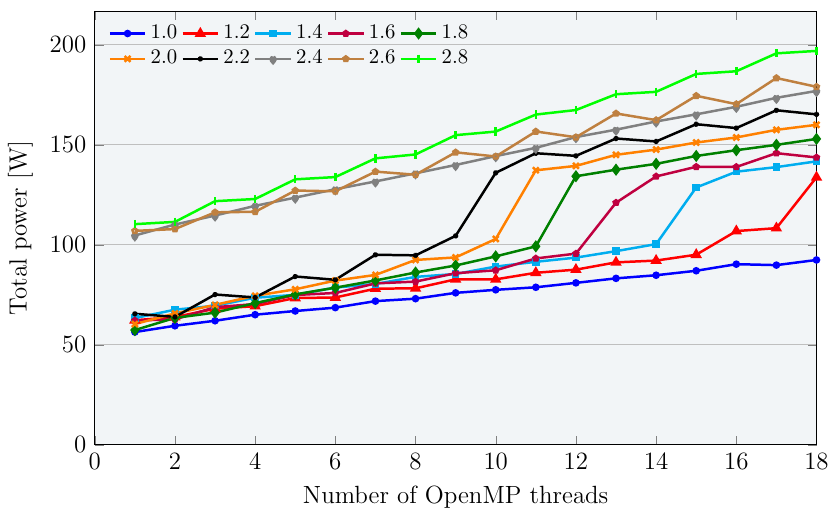}}
    \end{minipage}
    \caption{The impact of CPU frequency on (a, c) performance and (b, d) power characteristics for scalable \code{CalcKinematicsForElems} and memory-bound \code{CalcHourglassControlForElems} functions for OMP-ICL version.
    The CPU clock frequency is set at multiple values (1.0 -- 2.8 \GHZ), while uncore frequency remains unfixed, ranging from a minimum of 0.8 \GHZ~to a maximum of 2.5 \GHZ.
    }
    \label{fig:App-B}
\end{figure}
\begin{figure*}[b]
    \begin{minipage}{\textwidth}
        \centering
        \subfloat[ Total energy in OpenMP]{\includegraphics[scale=0.4]{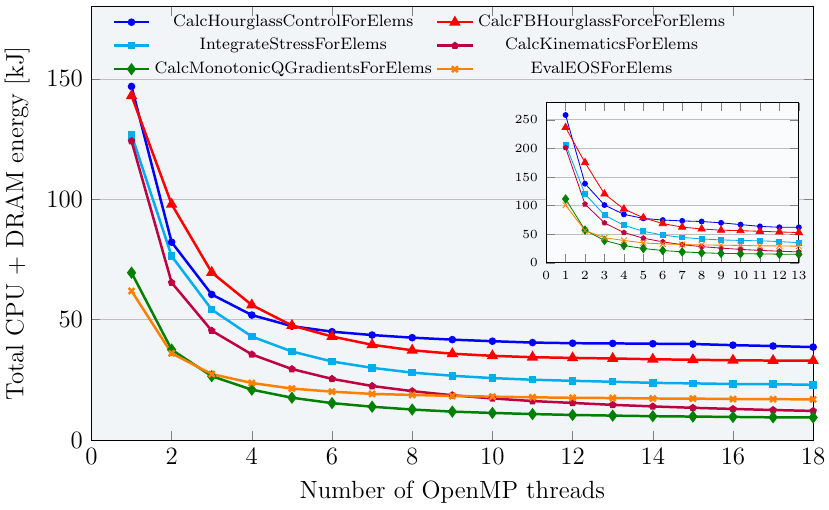}}\quad
        \subfloat[ Total energy in OpenMP] 
        {\includegraphics[scale=0.4]{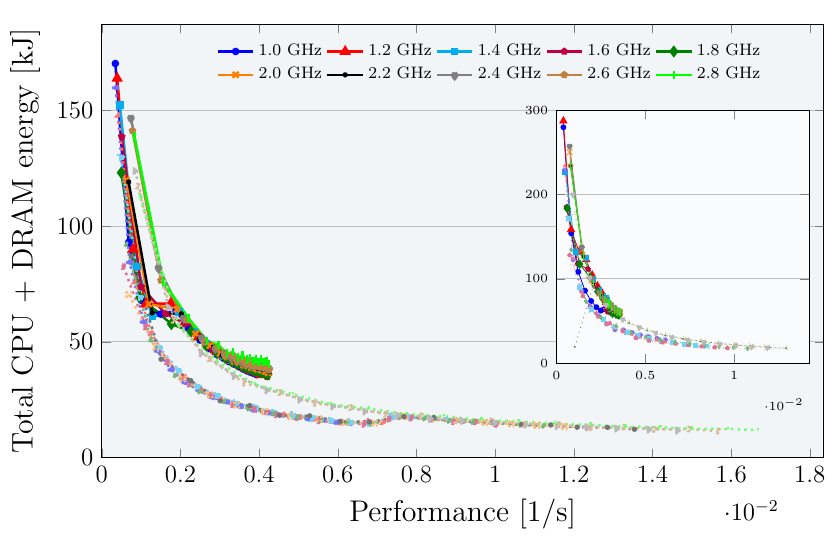}}\quad
        \subfloat[ Total energy in  MPI]{\includegraphics[scale=0.4]{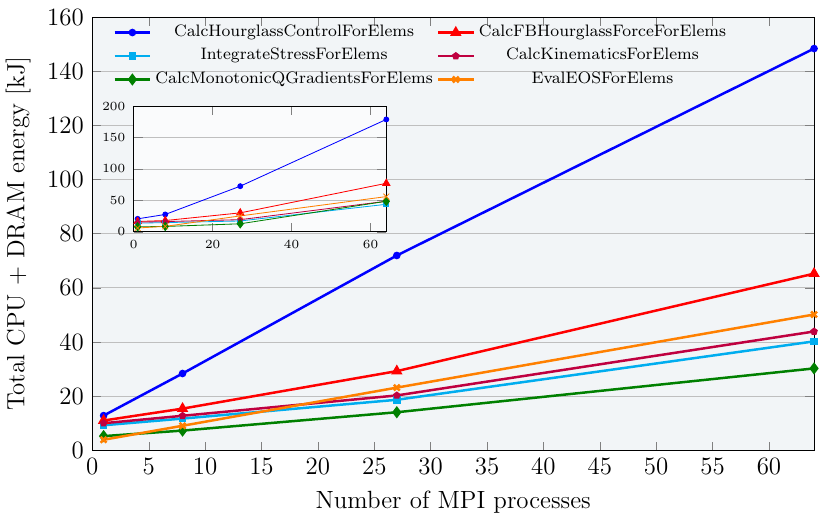}}
    \end{minipage}%
    \caption{Energy consumption of hot spot functions in (a, b) OpenMP and (c) MPI, with subplot view of ClusterB ccNUMA domain. 
    The second plot highlights the optimal EDP for different core counts and core frequencies (1.0 -- 2.8 \GHZ) at two corner cases: memory-bound hot spot function \code{CalcHourglassControlForElems} (lines) and non-memory-bound hot spot function \code{CalcKinematicsForElems} (dotted).
    The uncore frequency is not fixed, ranging from a minimum of 0.8 \GHZ~to a maximum of 2.5 \GHZ.}
    \label{fig:App-A}
\end{figure*}

\end{document}

%% file: setup.tex
\def\BibTeX{{\rm B\kern-.05em{\sc i\kern-.025em b}\kern-.08em
    T\kern-.1667em\lower.7ex\hbox{E}\kern-.125emX}}

\usepackage{amsmath,amsfonts}
\usepackage{algorithmic}
\usepackage{algorithm}
\usepackage{array}
\usepackage[caption=false,font=normalsize,labelfont=sf,textfont=sf]{subfig}
\usepackage{textcomp}
\usepackage{stfloats}
\usepackage{url}
\usepackage{verbatim}
\usepackage{graphicx}
\usepackage{cite}
\usepackage{balance}

\usepackage{acro}
\usepackage{adjustbox}
\usepackage{xtab,booktabs}
\usepackage{comment}
\usepackage{todonotes}
\usepackage{colortbl}
\usepackage{hyperref}
\usepackage{listings}
\usepackage[para,online,flushleft]{threeparttable}
\usepackage{longtable}
\usepackage{cite}
\usepackage{amssymb}
\usepackage{xcolor}
\usepackage{xspace}

\newcolumntype{B}{>{\columncolor{cyan!4!white}}p{0.46\textwidth}}
\newcolumntype{a}{>{\columncolor{gray!10!white}}c}
\newcolumntype{x}{>{\columncolor{green!13!white}}c}
\newcolumntype{y}{>{\columncolor{blue!13!white}}c}
\newcolumntype{z}{>{\columncolor{yellow!10!white}}c}
\newcolumntype{v}{>{\columncolor{red!10!white}}c}
\definecolor{OliveGreen}{rgb}{0,0.6,0}
\definecolor{ForestGreen}{RGB}{34,139,34}
\definecolor{myblue}{RGB}{37,165,203}
\definecolor{FAUblue}{rgb}{0.000, 0.2196, 0.3961}
\definecolor{myred}{RGB}{175,32,67}

\newcommand{\bq}{\begin{equation}}
\newcommand{\eq}{\end{equation}}

\newcommand{\second}{\mbox{s}}

\newcommand{\flop}{\mbox{flop}}

\newcommand{\bit}{\mbox{bit}}

\newcommand{\GBPS}{\mbox{G\bit/\second}}

\newcommand{\GBS}{\mbox{GB/\second}}

\newcommand{\GFS}{\mbox{G\flop/\second}}

\newcommand{\GHZ}{\mbox{GHz}}

\newcommand{\FB}{\mbox{F/B}}

\newcommand{\GB}{\mbox{GB}}

\newcommand{\GiB}{\mbox{GiB}}
\newcommand{\MiB}{\mbox{MiB}}
\newcommand{\KiB}{\mbox{KiB}}

\newcommand{\cma}{~,}

\newcommand{\LLC}{last-level cache\xspace}

\newcommand{\code}[1]{\texttt{#1}}

\lstdefinestyle{small}{
  language=C++,
  basicstyle=\fontsize{6.4}{6}\tt,
  showspaces=false,
  showstringspaces=false,
  breakindent=0pt,
  breaklines,
  captionpos=b,
  breakatwhitespace=true,
  numbers=left,
  numberstyle=\tiny,
  stepnumber=1,
  numbersep=7pt,
  linewidth=0.5\textwidth,
  xleftmargin=.02\textwidth, 
  xrightmargin=.02\textwidth,
  keywordstyle=\color{blue},
  tabsize=2,
  backgroundcolor=\color{white!97!black},
moredelim=**[is][\color{red}]{@}{@},
  moredelim=**[is][\color{green}]{|}{|},
}
\lstdefinestyle{smallWithColor}{
  basicstyle=\small\tt,
  language=C++,
  showspaces=false,
  showstringspaces=false,
  breakindent=0pt,
  breaklines,
  captionpos=b,
  breakatwhitespace=true,
  numbers=left,
  numberstyle=\tiny,
  stepnumber=1,
  numbersep=7pt,
  keywordstyle=\color{blue},
  tabsize=2,
  moredelim=**[is][\color{red}]{@}{@},
  moredelim=**[is][\color{green}]{|}{|},
  backgroundcolor=\color{white!97!black},
}

\lstdefinestyle{footnotesizeStyle}{
  basicstyle=\footnotesize\tt,
  language=C++,
  showspaces=false,
  showstringspaces=false,
  breakindent=0pt,
  breaklines,
  captionpos=b,
  breakatwhitespace=true,
  numbersep=5pt,
  keywordstyle=\color{blue},
  tabsize=2,
  moredelim=**[is][\color{red}]{@}{@},
  moredelim=**[is][\color{green}]{|}{|},
  backgroundcolor=\color{white!97!black},
}

\usepackage[htt]{hyphenat} 
  
\newcommand{\enquote}[1]{``#1''}  
\newcommand{\CODE}[1]{\texttt{#1}}

\colorlet{backgroundcol}{cyan!10!white}
\newcommand{\highlight}[1]{%
	\par\noindent
	\fcolorbox{black}{backgroundcol}{%
		\parbox{\dimexpr\linewidth-2\fboxsep\relax}{%
			#1
		}%
}}

\colorlet{backgroundcolumn}{yellow!10!white}

\newcommand{\acrodef}[2]{\DeclareAcronym{#1}{short={#1},long={#2}}}
\acrodef{CER}{commu\-ni\-cation-to-exe\-cu\-tion ratio}
\acrodef{ChebFD}{Chebyshev filter diagonalization}
\acrodef{EOS}{Equation of State}
\acrodef{HPCG}{High Performance Conjugate Gradient}
\acrodef{LBM}{Lattice Boltzmann Menthod}
\acrodef{LULESH}{Livermore Unstructured Lagrangian Explicit Shock Hydrodynamics}
\acrodef{MG}{multigrid}
\acrodef{MST}{MPI-augmented STREAM Triad}
\acrodef{SPEC}{Standard Performance Evaluation Corporation}
\acrodef{SoA}{structure of arrays}
\acrodef{SpMVM}{sparse matrix-vector multiplication}

\usepackage{pgfplots}
\usepackage{cancel}
\usepackage{pgfplotstable}
\usepackage{multirow}
\usepgfplotslibrary{colorbrewer}
\usetikzlibrary{pgfplots.statistics, pgfplots.colorbrewer} \usepackage{forest}
\usetikzlibrary{arrows.meta, shapes.geometric, calc, shadows}
\usetikzlibrary{fit,calc}
\usepackage{smartdiagram}

\colorlet{mygreen}{green!75!black}
\colorlet{col1in}{red!30}
\colorlet{col1out}{red!40}
\colorlet{col2in}{mygreen!40}
\colorlet{col2out}{mygreen!50}
\colorlet{col3in}{blue!30}
\colorlet{col3out}{blue!40}
\colorlet{col4in}{mygreen!20}
\colorlet{col4out}{mygreen!30}
\colorlet{col5in}{blue!10}
\colorlet{col5out}{blue!20}
\colorlet{col6in}{blue!20}
\colorlet{col6out}{blue!30}
\colorlet{col7out}{orange}
\colorlet{col7in}{orange!50}
\colorlet{col8out}{orange!40}
\colorlet{col8in}{orange!20}
\colorlet{linecol}{blue!60}
\usepgfplotslibrary{colorbrewer}

\usetikzlibrary{decorations.pathmorphing}
\tikzset{discont/.style={decoration={zigzag,segment length=1.5pt, amplitude=4pt},decorate}}
\def\discontarrow(#1)(#2)(#3)(#4);{
	\draw[discont] (#2) -- (#3);
	\draw[] (#1) -- (#2) (#3) -- (#4);
}

\usepgflibrary{decorations.fractals}
\pgfplotstableset{
rowfont/.style={
	postproc cell content/.append code={
		\count0=\pgfplotstablerow
		\advance\count0 by1
		\ifnum\count0=#1
		\pgfkeysalso{@cell content/.add={\bfseries\boldmath\color{red}}{}}
		\fi
	},
},
colfont/.style={
	postproc cell content/.append code={
		\count0=\pgfplotstablecol
		\advance\count0 by1
		\ifnum\count0=#1
		\pgfkeysalso{@cell content/.add={\color{black}}{}}
		\fi
	},
},
colfontred/.style={
	postproc cell content/.append code={
		\count0=\pgfplotstablecol
		\advance\count0 by1
		\ifnum\count0=#1
		\pgfkeysalso{@cell content/.add={\bfseries\boldmath\color{red}}{}}
		\fi
	},
},
colfontwhite/.style={
	postproc cell content/.append code={
		\count0=\pgfplotstablecol
		\advance\count0 by1
		\ifnum\count0=#1
		\pgfkeysalso{@cell content/.add={\bfseries\boldmath\color{white}}{}}
		\fi
	},
},	
}

\usepackage{colortbl}
\usepgfplotslibrary{colormaps}
\pgfplotsset{compat=1.17}
\pgfplotsset{colormap/Set3} 
\usepackage{cellspace}
\newcolumntype{C}{}
\newlength{\cellspacelimit}
\pgfplotstableset{
	/color cells/min/.initial=0,
	/color cells/max/.initial=100,
	/color cells/textcolor/.initial=,
	color cells/.code={%
		\pgfqkeys{/color cells}{#1}%
		\pgfkeysalso{%
			postproc cell content/.code={%
				\begingroup
				\pgfkeysgetvalue{/pgfplots/table/@preprocessed
					cell content}\value
				\ifx\value\empty
				\endgroup
				\else
				\pgfmathfloatparsenumber{\value}%
				\pgfmathfloattofixed{\pgfmathresult}%
				\let\value=\pgfmathresult
				\pgfplotscolormapaccess
				[\pgfkeysvalueof{/color cells/min}:\pgfkeysvalueof{/color
					cells/max}]
				{\value}
				{\pgfkeysvalueof{/pgfplots/colormap name}}%
				\pgfkeysgetvalue{/pgfplots/table/@cell content}\typesetvalue
				\pgfkeysgetvalue{/color cells/textcolor}\textcolorvalue
				\toks0=\expandafter{\typesetvalue}%
				\xdef\temp{%
					\noexpand\pgfkeysalso{%
						@cell content={%
							\noexpand\cellcolor[rgb]{\pgfmathresult}%
							\noexpand\definecolor{mapped
								color}{rgb}{\pgfmathresult}%
							\ifx\textcolorvalue\empty
							\else
							\noexpand\color{\textcolorvalue}%
							\fi
							\the\toks0 %
						}%
					}%
				}%
				\endgroup
				\temp
				\fi
			}%
		}%
	},
	every column/.style={column type={Cc!{\color{white}\vrule width 1pt}}},
	every first column/.style={reset styles, column type={l}, string type},
	every head row/.style={before row=\toprule, after row=\midrule},
	after row={\noalign{\global\arrayrulewidth=1pt}\arrayrulecolor{white}\hline},
	every last row/.style={after row=\arrayrulecolor{black}\bottomrule}, 
	@content options for rows/.style 2 args={
		postproc cell content/.add code={%
			\def\isInInputList{0}%
			\pgfplotsforeachungrouped \II in {#1} {%
				\ifnum\II<0
				\count0=\II
				\advance\count0 by\pgfplotstablerows
				\edef\II{\the\count0 }%
				\fi
				\ifnum\II=\pgfplotstablerow\relax
				\def\isInInputList{1}%
				\fi
			}%
			\if1\isInInputList%
			\pgfkeysalso{#2}%
			\fi
		}{},
	},
}

\pgfkeys{/pgf/number format/.cd,1000 sep={\,},fixed,precision=3}

\makeatletter
\pgfplotsset{
	boxplot/hide outliers/.code={
		\def\pgfplotsplothandlerboxplot@outlier{}%
	}
}
\makeatother

\usepgfplotslibrary{units}
\usetikzlibrary{arrows, calc, spy, patterns, backgrounds,decorations.pathreplacing, arrows.meta}

\pgfplotsset{
	my ybar legend/.style={
		legend image code/.code={
			\draw [##1] (0cm,-0.6ex) rectangle +(1.75em,1.1ex);
		},
	},
}

\makeatletter
\pgfdeclarepatternformonly[\hatchdistance,\hatchthickness]{flexible hatch}
{\pgfqpoint{0pt}{0pt}}
{\pgfqpoint{\hatchdistance}{\hatchdistance}}
{\pgfpoint{\hatchdistance-1pt}{\hatchdistance-1pt}}%
{
	\pgfsetcolor{\tikz@pattern@color}
	\pgfsetlinewidth{\hatchthickness}
	\pgfpathmoveto{\pgfqpoint{0pt}{0pt}}
	\pgfpathlineto{\pgfqpoint{\hatchdistance}{\hatchdistance}}
	\pgfusepath{stroke}
}

\makeatletter
\pgfplotsset{
	discontinuous line/.code={
		\pgfkeysalso{mesh, shorten <=#1, shorten >=#1,
			legend image code/.code={
				\draw [##1, shorten <=0cm] (0cm,0cm) -- (0.3cm,0cm);
				\draw [only marks] plot coordinates {(0.3cm,0cm)};
				\draw [##1, shorten >=0cm] (0.3cm,0cm) -- (0.6cm,0cm);
		}}
		\def\pgfplotsplothandlermesh@VISUALIZE@std@fill@andor@stroke{%
			\pgfplotspatchclass{\pgfplotsplothandlermesh@patchclass}{fill path}%
			\pgfplotsplothandlermesh@definecolor
			\pgfusepath{stroke}
			\pgfplotsplothandlermesh@show@normals@if@configured
		}%
	},
	discontinuous line/.default=1.5mm
}
\makeatother

\usetikzlibrary{patterns}
	\newlength{\hatchspread}
	\newlength{\hatchthickness}
	\newlength{\hatchshift}
	\newcommand{\hatchcolor}{}
	\tikzset{hatchspread/.code={\setlength{\hatchspread}{#1}},
		hatchthickness/.code={\setlength{\hatchthickness}{#1}},
		hatchshift/.code={\setlength{\hatchshift}{#1}},
		hatchcolor/.code={\renewcommand{\hatchcolor}{#1}}}
	\tikzset{hatchspread=3pt,
		hatchthickness=0.4pt,
		hatchshift=0pt,
		hatchcolor=black}
	\pgfdeclarepatternformonly[\hatchspread,\hatchthickness,\hatchshift,\hatchcolor]
	{custom north west lines}
	{\pgfqpoint{\dimexpr-2\hatchthickness}{\dimexpr-2\hatchthickness}}
	{\pgfqpoint{\dimexpr\hatchspread+2\hatchthickness}{\dimexpr\hatchspread+2\hatchthickness}}
	{\pgfqpoint{\dimexpr\hatchspread}{\dimexpr\hatchspread}}
	{
		\pgfsetlinewidth{\hatchthickness}
		\pgfpathmoveto{\pgfqpoint{0pt}{\dimexpr\hatchspread+\hatchshift}}
		\pgfpathlineto{\pgfqpoint{\dimexpr\hatchspread+0.15pt+\hatchshift}{-0.15pt}}
		\ifdim \hatchshift > 0pt
		\pgfpathmoveto{\pgfqpoint{0pt}{\hatchshift}}
		\pgfpathlineto{\pgfqpoint{\dimexpr0.15pt+\hatchshift}{-0.15pt}}
		\fi
		\pgfsetstrokecolor{\hatchcolor}
		\pgfusepath{stroke}
	}
	\pgfdeclarepatternformonly[\hatchspread,\hatchthickness,\hatchshift,\hatchcolor]
	{custom north east lines}
	{\pgfqpoint{\dimexpr-2\hatchthickness}{\dimexpr-2\hatchthickness}}
	{\pgfqpoint{\dimexpr\hatchspread+2\hatchthickness}{\dimexpr\hatchspread+2\hatchthickness}}
	{\pgfqpoint{\dimexpr\hatchspread}{\dimexpr\hatchspread}}
	{
		\pgfsetlinewidth{\hatchthickness}
		\pgfpathmoveto{\pgfqpoint{\dimexpr\hatchshift-0.15pt}{-0.15pt}}
		\pgfpathlineto{\pgfqpoint{\dimexpr\hatchspread+0.15pt}{\dimexpr\hatchspread-\hatchshift+0.15pt}}
		\ifdim \hatchshift > 0pt
		\pgfpathmoveto{\pgfqpoint{-0.15pt}{\dimexpr\hatchspread-\hatchshift-0.15pt}}
		\pgfpathlineto{\pgfqpoint{\dimexpr\hatchshift+0.15pt}{\dimexpr\hatchspread+0.15pt}}
		\fi
		\pgfsetstrokecolor{\hatchcolor}
		\pgfusepath{stroke}
	}

\usepackage{awesomebox}
\usepackage{lipsum}
\usepackage{tikz}
\tikzset{pics/speedometer/.style={code={
 \foreach \X/\Y [count=\Z] in {green!70!black/low,orange/medium,red/high}
  {\fill[fill=\X] (240-\Z*60:4) arc(240-\Z*60:180-\Z*60:4) -- 
    (180-\Z*60:3) arc(180-\Z*60:240-\Z*60:3) -- cycle;}
   \fill (180-#1+8:0.3) arc (180-#1+8:180-#1+344:0.3) -- (180-#1-0.5:3.25)
   -- (180-#1+0.5:3.25) --  cycle;
  }}}
\newsavebox\ZeroSpeed  
\newsavebox\LowSpeed  
\newsavebox\MediumSpeed  
\newsavebox\HighSpeed  
\sbox\ZeroSpeed{\scalebox{0.07}{\tikz{\pic{speedometer=0};}}}
\sbox\LowSpeed{\scalebox{0.07}{\tikz{\pic{speedometer=45};}}}
\sbox\MediumSpeed{\scalebox{0.07}{\tikz{\pic{speedometer=90};}}}
\sbox\HighSpeed{\scalebox{0.07}{\tikz{\pic{speedometer=135};}}}

%% file: figures/tab_systems.tex
	\begin{table}[t]
		\centering
		\caption{Key hardware and software attributes of systems.} 
		\label{tab:systems}
		\begin{adjustbox}{width=0.48\textwidth}
				\setlength\extrarowheight{-0.7pt}
                \setlength\tabcolsep{2pt}
                \Huge
                \begin{tabular}[fragile]{c>{~}lyz}
                	\toprule
                	\rowcolor[gray]{0.9}
                	\cellcolor[gray]{0.9}&Systems  &  ClusterA & ClusterB \\
                	\midrule
                	\cellcolor[gray]{0.9}& Processor   & Intel Xeon Ice Lake  & Intel Xeon Sapphire Rapids \\    
                	\cellcolor[gray]{0.9}&Processor Model       & \href{https://ark.intel.com/content/www/us/en/ark/products/212459/intel-xeon-platinum-8360y-processor-54m-cache-2-40-ghz.html}{Platinum 8360Y}  & \href{https://ark.intel.com/content/www/us/en/ark/products/231728/intel-xeon-platinum-8470-processor-105m-cache-2-00-ghz.html}{Platinum 8470}   \\
                	\cellcolor[gray]{0.9}&Base clock speed  & $2.4$~\GHZ\  &  $2.0$~\GHZ\ \\
                	\cellcolor[gray]{0.9}&Physical cores per node     & 72    & 104      \\
                	\cellcolor[gray]{0.9}&ccNUMA domains per node   & 4  & 8   \\
                	\cellcolor[gray]{0.9}&Sockets per node   & 2  & 2   \\
                	\cellcolor[gray]{0.9}&Per-core L1 and L2 cache sizes &  $48$~\KiB\ (L1) + $1.25$~\MiB\ (L2)  & $48$~\KiB\ (L1) + $2$~\MiB\ (L2) \\
                	\cellcolor[gray]{0.9}&Shared \LLC size &  $54$~\MiB\ (L3)  & $105$~\MiB\ (L3) \\
                	\cellcolor[gray]{0.9}&Memory per node & $4\times 64$~\GiB\  & $8\times 128$~\GiB\ \\
                	\cellcolor[gray]{0.9}&Socket memory type & 8 channels DDR4-3200  & 8 channels DDR5-4800 \\
                	\cellcolor[gray]{0.9}&Theoretical socket memory bandwidth &  $2\times 102.4$~\GBS\ & $4\times 76.8$~\GBS\ \\
                    \cellcolor[gray]{0.9} & Socket thermal design power & 250 W & 350 W\\
                    \multirow{-11}{*}{\rotatebox{90}{\cellcolor[gray]{0.9} Micro-architecture}} & No. of Ultra Path Interconnect links & 3 & 4 \\
                    \midrule
                	\cellcolor[gray]{0.9}&Node interconnect     & HDR100 Infiniband  & HDR100 Infiniband    \\
                	\cellcolor[gray]{0.9}&Interconnect topology  & Fat-tree  & Fat-tree \\
                	\cellcolor[gray]{0.9}&Raw bandwidth per link \& direction &  $100$~\GBPS\ & $100$~\GBPS\   \\
                    \cellcolor[gray]{0.9}&Parallel filesystem (capacity) & Lustre-based ($3.5$~PB) & Lustre-based ($3.5$~PB)\\ 
                    \multirow{-5}{*}{\rotatebox{90}{\cellcolor[gray]{0.9} Network}}&Aggregated parallel I/O bandwidth & $>20$~\GBS\ & $>20$~\GBS\ \\
                	
                	\midrule
                	\cellcolor[gray]{0.9}&Compiler    & Intel v2023up2.1   & Intel v2023up2.1 \\
                	\cellcolor[gray]{0.9}&Optimization flags & -O3 -qopt-zmm-usage=high & -O3 -qopt-zmm-usage=high \\
                	\cellcolor[gray]{0.9}&SIMD &-xCORE-AVX512 &  -xCORE-AVX512\\
                	\cellcolor[gray]{0.9}&Message passing library  & Intel \verb.MPI. v2021u10    &  Intel \verb.MPI. v2021u10 \\
                	\multirow{-5}{*}{\rotatebox{90}{\cellcolor[gray]{0.9} Software}}&Operating system    &  AlmaLinux release 8.10  &  AlmaLinux release 8.10 \\
                	
                	\midrule
                	\cellcolor[gray]{0.9}& Intel \CODE{VTune} version   &   2023.2.0    &  2023.2.0  \\
                	\cellcolor[gray]{0.9}&\CODE{LIKWID} version & 5.3.0  & 5.3.0-spr\\
                	\multirow{-3}{*}{\rotatebox{90}{\cellcolor[gray]{0.9} Tools}}&\CODE{LIKWID} flags  & -g MEM\_DP -m    & -g MEM\_DP -m \\
                	\bottomrule
                \end{tabular}
		\end{adjustbox}
\medskip
	\end{table}

%% file: figures/tab_algoritm.tex
	\begin{table}[t]
		\centering
		\caption{Algorithm in a single LULESH iteration executed in two steps: nodal updates (blue) and element updates (green) with communication (red) involved.} 
		\label{tab:Fig_LULESHAlgo}
		\begin{adjustbox}{width=0.48\textwidth}
				\setlength\extrarowheight{-0.7pt}
                \setlength\tabcolsep{2pt}
                \Huge
                \begin{tabular}[fragile]{yx}
                	\toprule
                	\rowcolor[gray]{0.9} Nodal updates & Element updates \\
                	\midrule
                	Calculate forces using stress values  & Calculate kinematics \\ 
                 Perform hourglass correction &  Calculate artificial viscosity for \\ 
                 &  single- and multi-material problems\\ 
                 Calculate acceleration (using appropriate  & \cellcolor{red!20} Communicate viscosity gradients \\ 
                 symmetry boundary conditions) &  \\
                 Calculate velocity &  Apply material properties\\ 
                 Update positions &  \\ 
                 \cellcolor{red!20} Communicate positions&  \\ 
                	\bottomrule
                \end{tabular}
		\end{adjustbox}
   \medskip
	\end{table}

%% file: figures/Kernels/tab_Roofline.tex
\begin{table*}[t]
    \centering
	\caption{Single-threaded predicted and measured computational intensities (Eq \ref{eq:I}) in~[\FB] for the five memory-bound hot spots.
    Multi-threaded measured intensity values, scaled to the entire ccNUMA domain in OpenMP
    , are given in parentheses.
    } 
	\label{tab:Roofline}
	\begin{adjustbox}{width=0.98\textwidth}
            \setlength\extrarowheight{4pt} 
            \setlength\tabcolsep{3pt} 
            \begin{threeparttable}
            \begin{tabular}[fragile]{lccvzxy}
            	\toprule
                \rowcolor[gray]{0.9}
            \multicolumn{1}{c|}{} & \multicolumn{2}{c|}{Predicted computational intensities } & \multicolumn{4}{c}{Measured computational intensities} \\ 
            	\rowcolor[gray]{0.9}
                \multicolumn{1}{c|}{Hot spot} & &  \multicolumn{1}{c|}{} & \mbox{OMP-ICL}  & \mbox{OMP-SPR} & \mbox{MPI-ICL}\mbox{$^\star$} & \mbox{MPI-SPR}\mbox{$^\star$} \\
                \rowcolor[gray]{0.9} \multicolumn{1}{c|}{} & \mbox{single-threaded} & \multicolumn{1}{c|}{(\mbox{multi-threaded}\mbox{$^\diamond$})}  & \mbox{single-threaded (multi-threaded)} & \mbox{single-threaded (multi-threaded)} & \mbox{single-threaded } & \mbox{single-threaded }\\
            	\midrule
                \cellcolor[gray]{0.9} First & $\frac{325}{636 + 584} = 0.27$ &  ( -- ) & $\frac{325}{633 + 532} = 0.28~(\frac{325}{556 + 738} = 0.25) $   & $\frac{325}{591 + 490} = 0.3~(\frac{325}{500 + 730} =0.26) $ & $\frac{325}{633 + 532} = 0.28$ & $\frac{325}{593 + 492} = 0.3$ \\                
                \cellcolor[gray]{0.9} Second & $\frac{824}{584 + 48}=1.39$  & ($\frac{824}{776 + 240}=0.8$) & $\frac{824}{547 + 48} = 1.39~(\frac{824}{730 + 368}=0.75)$   & $\frac{824}{541 + 53} = 1.39~(\frac{824}{716 + 354}=0.77) $ & $\frac{824}{544 + 48} = 1.39$ & $\frac{824}{540 + 52} = 1.39$ \\

                \cellcolor[gray]{0.9} Third & $\frac{386}{144+56} = 2.01$  & ($\frac{386}{336+248} = 0.66$)
 & $\frac{386}{139 + 48} = 2.07~(\frac{386}{374 + 367}=0.52)$   & $\frac{386}{135 + 49} = 2.09~(\frac{386}{354 + 357}=0.54)$ & $\frac{386}{138 + 48} = 2.08\mbox{$^\mathsection$}$ & $\frac{386}{134 + 48} = 2.12\mbox{$^\mathsection$}$ \\

               \cellcolor[gray]{0.9} Fourth & $\frac{205}{240 + 56} = 0.69$   & ( -- ) & $\frac{205}{214 + 83} = 0.69~(\frac{205}{207 + 95}=0.68)$   & $\frac{205}{216 + 88} = 0.67~(\frac{205}{205 + 94}=0.68)$ & $\frac{205}{213 + 83} = 0.69$ & $\frac{205}{215 + 88} = 0.67$ \\

                \cellcolor[gray]{0.9} Fifth & $\frac{6}{48 + 16} = 0.1$  & ( -- ) & $\frac{6}{45 + 14} = 0.1~(\frac{6}{37 + 18}=0.1)$   & $\frac{6}{42 + 15} = 0.1~(\frac{6}{38 + 16}=0.1)$ & $\frac{6}{45 + 14} = 0.1$ & $\frac{6}{39 + 14} = 0.1$ \\
            	\bottomrule
            \end{tabular}
            \begin{tablenotes}
        \footnotesize
        \item \mbox{$^\diamond$} Code modifications specific to the multi-threaded case are included in the prediction wherever applicable.

        \item \mbox{$^\star$} The multi-threaded values for MPI were unnecessary and, therefore, omitted. 
        
        \item \mbox{$^\mathsection$} This \code{IntegrateStressForElems} function is the hot spot that becomes non-memory-bound in the MPI-parallelized implementation of LULESH.
        
        \noindent\rule[0.5ex]{\linewidth}{1pt}
    \end{tablenotes}
    \end{threeparttable}
	\end{adjustbox}
\end{table*}

%% file: figures/FullApp/tab_vectorization.tex
\begin{table}[t]
    \centering
	\caption{Scalar and vectorized double instructions for code variants. 
    } 
	\label{tab:vectorization}
	\begin{adjustbox}{width=0.48\textwidth}
            \setlength\extrarowheight{-0.7pt}
            \setlength\tabcolsep{2pt}
            \begin{tabular}[fragile]{lvzxy}
            	\toprule
            	\rowcolor[gray]{0.9}
                Version (column), Instructions (row) & scalar & 128-packed  & 256-packed & 512-packed \\
            	\midrule
                \cellcolor[gray]{0.9}OpenMP (ICL / SPR) & 93.9 & 1.1   & 4.7 & 0.3 \\
                \cellcolor[gray]{0.9}MPI (ICL / SPR) & 93.5 & 0 & 4.9 & 1.6  \\
            	\bottomrule
            \end{tabular}
	\end{adjustbox}
    \vspace{1.5em}
\end{table}